\documentclass[pre,twocolumn]{revtex4}

\usepackage{float}
\usepackage{setspace}
\usepackage{graphicx}        
\begin{document}

\title{Flow of power-law fluids in self-affine fracture channels}

\author{ Yiguang Yan$^{1,2}$ and Joel Koplik$^{1,3}$ }
\affiliation{Benjamin Levich Institute$^{1}$ and Departments of Mechanical
Engineering$^{2}$ and Physics$^{3}$,
City College of the City University of New York, New York, NY, 10031}

\date{\today}

\begin{abstract}
The two-dimensional pressure driven flow of non-Newtonian power-law fluids in
self-affine fracture channels at finite Reynolds number is calculated. The
channels have constant mean aperture and two values $\zeta$=0.5 and 0.8 of the Hurst
exponent are considered. The calculation is based on the lattice-Boltzmann method, using
a novel method to obtain a power-law variation in viscosity, and the behavior of
shear-thinning, Newtonian and shear-thickening liquids is compared.  Local aspects of
the flow fields, such as maximum velocity and pressure fluctuations, were studied, and
the non-Newtonian fluids were compared to the (previously-studied)  newtonian case.
The permeability results may be collapsed into a master curve of friction factor vs.
Reynolds number using a scaling similar to that employed for porous media flow,
and exhibits a transition from a linear regime to a more rapid variation at Re
increases.
\end{abstract}

\maketitle

\section{Introduction}
An understanding of flow and transport processes in geologically disordered
media is necessary for the efficient extraction of fluids from underground
hydrocarbon reservoirs. Situations where flow proceeds through networks of
connected fractures are particularly attractive, because the throughput is
generally much higher than may be achieved through intergranular porosity alone
\cite{adler,berkowitz,dietrich,sahimi}.  An important feature of subsurface 
fractures, which considerably
complicates the problem, is that the surfaces of naturally fractures rocks are
not smooth or even randomly rough, but rather are highly correlated self-affine
fractals \cite{bouchaud}. A second complication in the analysis is that typical
reservoir fluids are often complicated mixtures, which exhibit non-Newtonian
flow behaviors such as shear-thinning or shear-thickening.  Yet a third
difficulty is that the subsurface fracture flow often involves much higher
velocities than in the intergranular case, and the common simplification of
low-Reynolds number linear flow is inapplicable.

In this paper we use lattice Boltzmann calculations to elucidate the combined
effects of self-affinity, non-linear rheology and finite inertia in fluid flow
through a single fracture.  Previous authors have considered subsets of these
complications, but not all three simultaneously.  The flow of Newtonian fluids
in self-affine fractures at both low
\cite{drazer2000:permeability,drazer2002:transport} and finite \cite{skjetne1999:flow} 
$Re$ has an extensive literature.
Some controlled experiments on shear-thinning fluids in self-affine fractures at
low $Re$ have been reported \cite{auradou}. Lastly, experiments and
phenomenological models for non-linear fluid motion in intergranular porous
media at various $Re$ are available \cite{chhabra2001:review}. We anticipate
that flow in a fracture can be characterized in a manner similar to the latter
problem, since in both cases the key effect is that the random solid boundary of
the flow domain causes streamlines to wind around. One simplification which we
{\em can} exploit, however, is to focus on two-dimensional flows.  It is well
known that the flow of a single fluid in a straight channel differs only in
detail between two and three-dimensional cases, and furthermore, in porous media
flow in the analogous intergranular case, one sees the same flow laws for both
two and three dimensional geometries.

The approach taken in the paper follows the lines of our previous studies of
permeability \cite{drazer2000:permeability} and transport
\cite{drazer2002:transport} in self-affine fractures based on the
lattice-Boltzmann method, along with a procedure for incorporating power-law
viscosity variation similar to that developed previously \cite{drazer2005:nn}.
The discussion of inertial effects is influenced by previous studies for the
case of a Newtonian fluid in intergranular porous media
\cite{skjetne1999:flow}.
The fracture surface is generated numerically by a Fourier transform algorithm
and discretized on the regular lattice used in the flow problem.  The upper and
lower fracture surfaces bound the allowed nodes in the flow domain, a bounce-back 
condition enforces the no-slip boundary condition, and constant forcing
provides a pressure driven flow.  For power-law fluids, the lattice-Boltzmann
relaxation time is adjusted locally in space and time to provide the desired
relation between stress and strain.  The relation between imposed pressure drop
and total fluid flux provides the permeability, and the local flow fields are
analyzed to discuss the velocity, pressure and shear stress variations.
Some background on the flow geometry and calculational method is presented in
Section~\ref{sec:background}, the local analysis of the flow fields in
Sec.~\ref{sec:local}, the discussion of permeability is in Sec.~\ref{sec:perm},
and we summarize in Sec.~\ref{sec:concl}.

\section{Background}
\label{sec:background}

\subsection{Self-affine roughness}

In this subsection we review the characterization of self-affine fractures and
their numerical implementation. We consider a fracture surface without
overhangs, {\em i.e.}, the surface height $h(x,y)$ is a single-valued function
of the two coordinates $\mathbf{r}=(x,y)$ lying in the mean plane of the
surface. A {\em self-affine} fractal surface is one which displays different
scaling along the different spatial directions\cite{feder}, a statistical
self-similarity under the transformation
\begin{equation}
x \rightarrow \lambda x \mbox{ and } y \rightarrow \lambda y \Rightarrow
h(\mathbf{r}) \rightarrow \lambda^{\zeta} h( \mathbf{r} )
\end{equation}
where $\zeta$ is the Hurst or roughness exponent. Observations of a variety of
naturally fractured rock surfaces in different fracture modes yield just two
common values of $\zeta$, approximately 0.5 and 0.8. We further assume that the
surface has spatial isotropy in its mean plane. The surface is further
characterized by the amplitude of the roughness, or equivalently the prefactor
$C_0$ in the height-height correlation function,
\begin{equation}
\langle [h(\mathbf{r+\Delta})- h(\mathbf{r})]^2\rangle = C_0
(|\mathbf{\Delta}|/\ell)^{2\zeta}
\label{eq:cf}
\end{equation}
where the intrinsic length scale $\ell$ might be the grain size in experiment or
the lattice spacing in a calculation. In practice we generate self-affine
surfaces using the a Fourier synthesis method \cite{voss} as in
\cite{drazer2000:permeability}.

\begin{figure}[htb]
\begin{center}
\includegraphics[width=0.8\linewidth]{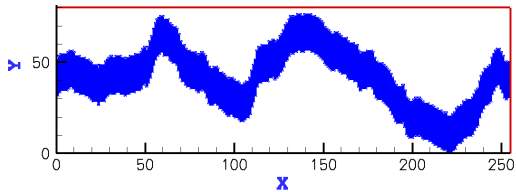}
\caption{Geometry of a typical self-affine fracture composed of two
complimentary self-affine surfaces with $\zeta$=0.8. \label{fig:saff-geom}}
\end{center}
\end{figure}

A self-affine fracture channel is made of two complementary self-affine surfaces
separated by a gap, and in some cases the surfaces are shifted relative to each
other parallel to the mean plane. The statistical properties of the fracture are
specified by the Hurst exponent, the mean aperture between two surfaces, the
shift distance, if any, and by the amplitude of the roughness. The height
fluctuations of a single self-affine surface increase with its lateral extent
$L$, so that the difference between the maximum and minimum heights scales as
$(L/\ell)^\zeta$, and we consider the limit $H\ll R < L$, as shown for a typical
fracture in Fig.~\ref{fig:saff-geom}.

Note that the effective flow diameter of the fracture varies along its length
and can be much smaller than the mean aperture, due to the tortuosity of the
channel.  When a lateral shift is present, the aperture varies locally as well, and
furthermore if $H$ is too small the sides of the fracture may overlap.

\subsection{The lattice-Boltzmann method}
Since the flow domain is bounded by highly irregular surfaces, the lattice Boltzmann
method \cite{succi} is particularly convenient for fluid mechanical calculations, since
the excluded solid region may be simply specified by a mask.
If $f_i(\mathbf{x},t)$ is the velocity distribution function (VDF) for particles moving
in direction $i$ at lattice site $\mathbf{x}$ at time $t$, then the discrete Boltzmann
equation which evolves the distribution is
\begin{equation}
f_i(\mathbf{x}+\mathbf{e}_i, t+1)
=f_i(\mathbf{x},t)+\Omega_i(f(\mathbf{x},t)),
\label{eq:lbm}
\end{equation}
Here the $\mathbf{e}_i$ are unit lattice vectors, the lattice
spacing and the time step are both set equal to one,
$\Omega_i(f(\mathbf{x},t))$ is collision operator which
redistributes the VDF along different directions, and the spatial and temporal step 
discretizes in single unit. To
recover the Navier-Stokes equation of fluid flow starting from
Boltzmann equation, moments of the VDF satisfy the constraints
\begin{equation}
 \rho=\Sigma_i f_i \quad  \rho\mathbf{u}=\sum_i f_i \mathbf{e}_i \quad
\mbox{\boldmath $\sigma$} = -\rho c_s^2\,\mathbf{I} - 
(1-{1\over 2\tau})\sum_i f_i \mathbf{e}_i \mathbf{e}_i
\end{equation}
which relate the distribution function to the continuum density, velocity and stress fields,
and where $c_s$ is the sound speed.  The collision operator is treated in the
BGK approximation using a single characteristic relaxation time $\tau$,
\begin{equation}
 \Omega_i(f(\mathbf{x},t))=-\frac{1}{\tau}(f_i(\mathbf{x},t)-
f_i^{eq}(\mathbf{x},t)),
 \label{eq:bgk}
\end{equation}
where $f_i^{eq}(\mathbf{x},t)$ is equilibrium distribution function. The
relaxation time $\tau$ is related to the kinematic viscosity of the fluid  by
$\nu=(2\tau-1)/6$. To simulate a constant pressure gradient we add a
a constant body force term to the right hand side of eq.\ref{eq:lbm}.
equation eq.\ref{eq:lbm} and \ref{eq:bgk} to Navier-Stokes equation.
More details may be found in \cite{succi}, and a recent review of flow simulations in
this context is presented by Verberg and Ladd\cite{verberg2001:lbmreview}.

\subsection{Power-law fluids}
The basic idea in extending the lattice Boltzmann method to
power-law fluids was presented by Aharonov and Rothman
\cite{rothman1993:nonNewtonian}, and consists of adjusting the
relaxation time $\tau$ {\em locally} so as  to achieve the desired
ratio of stress to strain rate. Here we consider power-law fluids
using a generalized Newtonian model, as in
\cite{drazer2005:nn}, where the relation between the stress
tensor $\sigma_{\alpha\beta}$ and the strain rate tensor
$D_{\alpha\beta}=1/2((\partial_\beta u_\alpha+\partial_\alpha
u_\beta)$ is similar to that for Newtonian fluids,
$\sigma_{\alpha\beta}=2\mu D_{\alpha\beta}$, but the local viscosity
$\mu$ is a function of the invariants of the strain rate tensor. We
consider power-law fluids, $\mu=m \dot{\gamma}^{n-1}$, where the
case $0<n<1$ corresponds to shear-thinning,  $n>1$ corresponds to
shear-thickening, and $n=1$ recovers linear Newtonian fluids, where
the local shear rate $\dot\gamma$ is related to the second invariant
of $D_{ij}$ via $\dot\gamma=\left(2\mathbf{D:D}\right)^{1/2}$. The
procedure in \cite{drazer2005:nn} was to obtain the strain
rate tensor by numerical differentiation of the previously
calculated velocity field, then determine the appropriate local
viscosity and thence the local relaxation time. Here we adopt a
different procedure:  in the lattice Boltzmann method the strain
rate tensor is directly related to the velocity distribution
function by \cite{chen1998:lbm-review}
\begin{equation}
D_{\alpha\beta}= -
\frac{3}{2\rho\tau}\sum_i{(f_i- f_i^{eq})\mathbf{e}_{i\alpha}\mathbf{e}_{i\beta}};
\end{equation}
which should in turn equal $\sigma_{\alpha\beta}/2\mu$, there is a constraint on the
$f_i$ which is solved by iteration.

To validate the formulation of power-law fluids given above, we calculate the
velocity profile for pressure-driven flow in a smooth-walled channel of constant
aperture (a Hele-Shaw cell), which may be compared to an analytic solution of the
Navier-Stokes equation. Applying a pressure gradient $\Delta P/L=-G$ in the
$x$-direction, the velocity for a power-law fluid with rheological parameters $m,n>0$
as above in a channel of width $H$ is
\begin{equation}
    u_x(y)=\frac{n}{n+1}\left(\frac{G}{m}\right)^{1/n} \left(
\left|\frac{H}{2}\right|^{(n+1)/n}-\left| \frac{H}{2}-y\right|^{(n+1)/n} \right),
    \label{eqn:vpower-law}
\end{equation}
We also record the mean velocity $\overline{u}$ and the fluid flux $Q$
(per unit length in the passive third direction), which will be useful below:
\begin{equation}
Q = H\overline{u}=\int_0^H dy\, u_x(y) =
{n\over 2n+1}\left({H^2\over 2}\right)\left(\frac{GH}{2m}\right)^{1/n}
    \label{eq:flux}
\end{equation}
In the simulation, we begin with zero velocity and integrate Eq. (\ref{eq:lbm})
to steady state, using the convergence criterion
\begin{equation}
    \epsilon=\sum_x \frac{\Vert u(\mathbf{x},t)-u(\mathbf{x},t-1) \Vert}{\Vert
u(\mathbf{x},t)\Vert}<1.0\times10^{-6},
\end{equation}

\begin{figure}[htb]
\begin{center}
\includegraphics[scale=.2,angle=-90.]{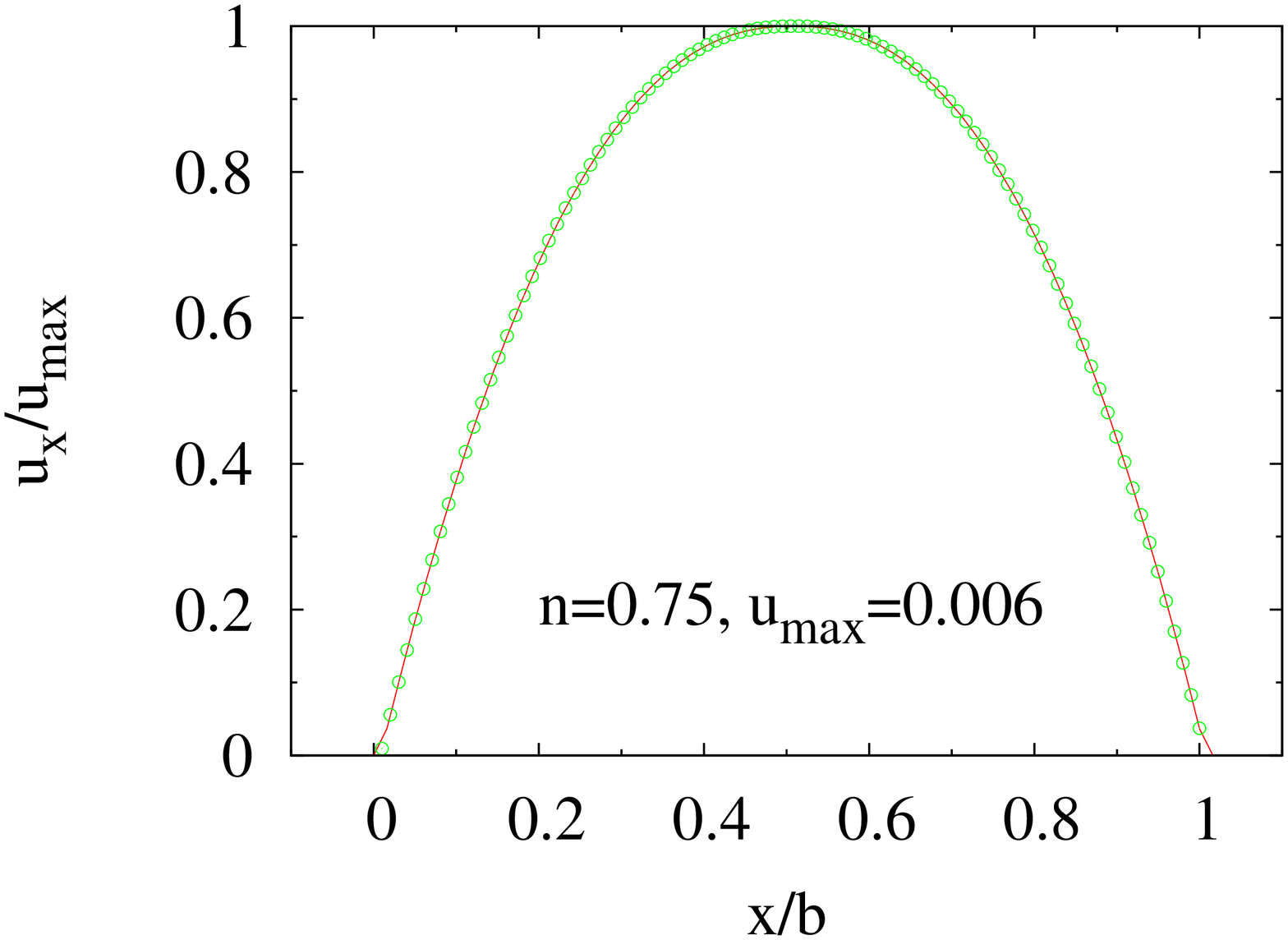}
\includegraphics[scale=.2,angle=-90.]{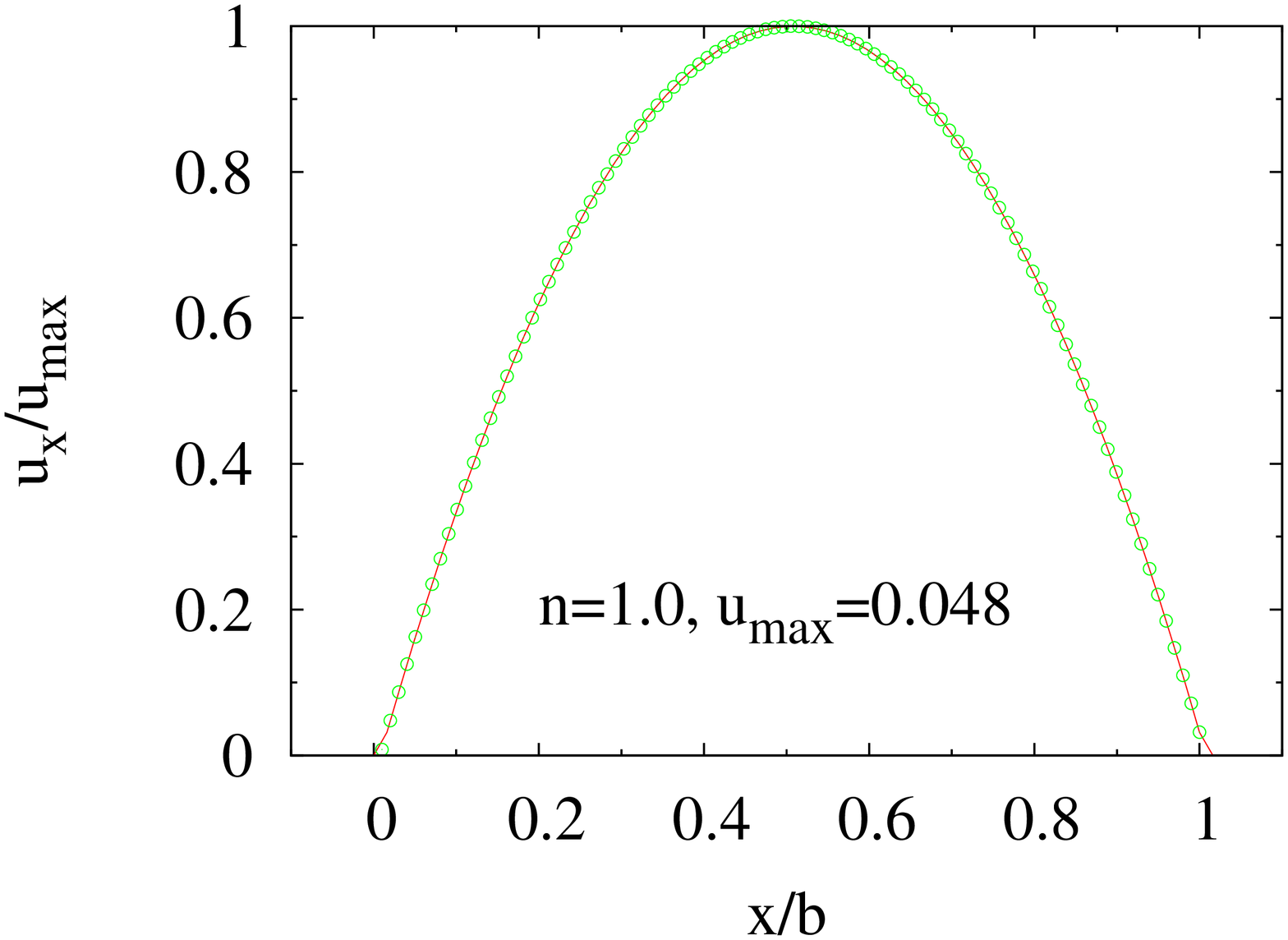}
\includegraphics[scale=.2,angle=-90.]{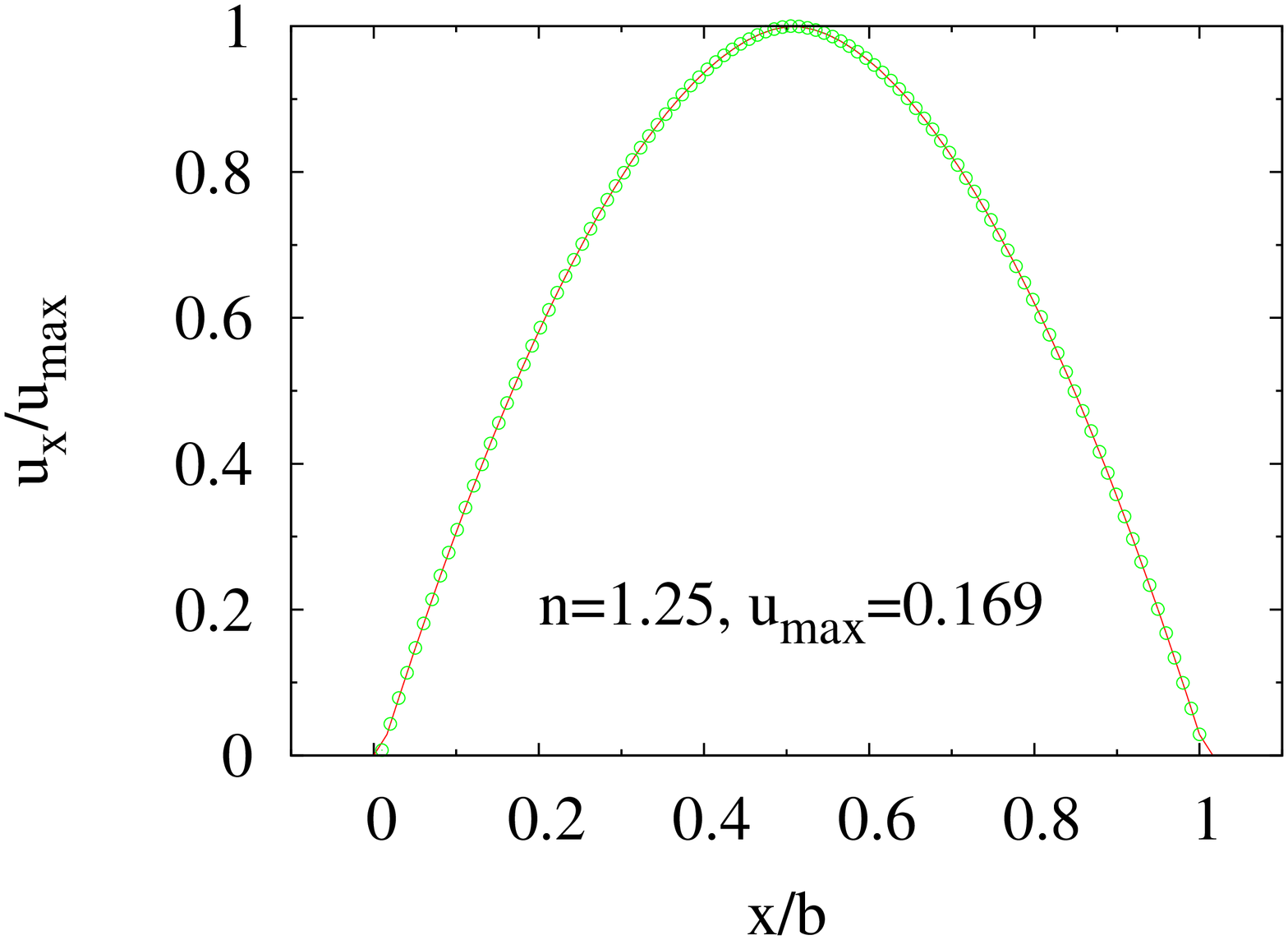}
\caption{Velocity profiles of power-law fluids with $m=0.01,n=0.75,1.0,1.25$
in a Hele-Shaw cell with pressure gradient $G=1\times 10^{-6}$.
The points are simulation results while the solid lines are the analytical solution
in (\ref{eqn:vpower-law}).  The maximum velocities for the three fluids are
$u_{max}=0.006,0.048,0.169$, respectively.}
\label{fig:vvalid-n75}
\end{center}
\end{figure}

For power-law indices $n$=0.75, 1.0 and 1.25, $m$=0.01, and pressure
gradient $G=1\times 10^{-6}$ we obtain the profiles shown in
Fig.~\ref{fig:vvalid-n75}, which agree with theory. In practice, as
with any numerical method, computational instabilities may occur for
substantially different values of the pressure gradient and fluid
index, but the algorithm could be extended there using techniques
such as multi-time step relaxation for the local shear viscosity
\cite{sullivan2006:porous}.

\section{Local analysis of the flow field}
\label{sec:local}
\begin{figure}[htb]
\begin{center}
\includegraphics[scale=.15]{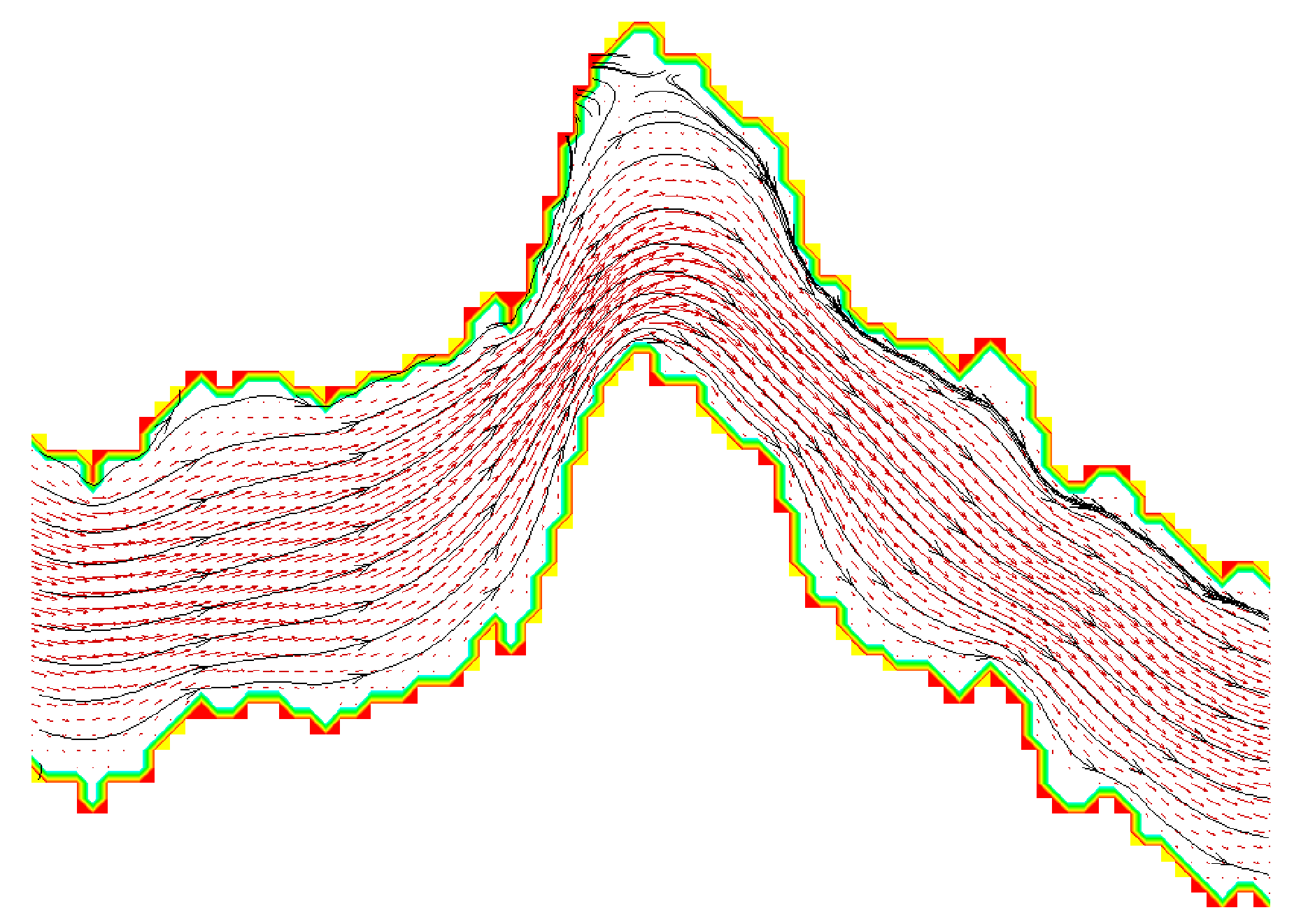}
\includegraphics[scale=.15]{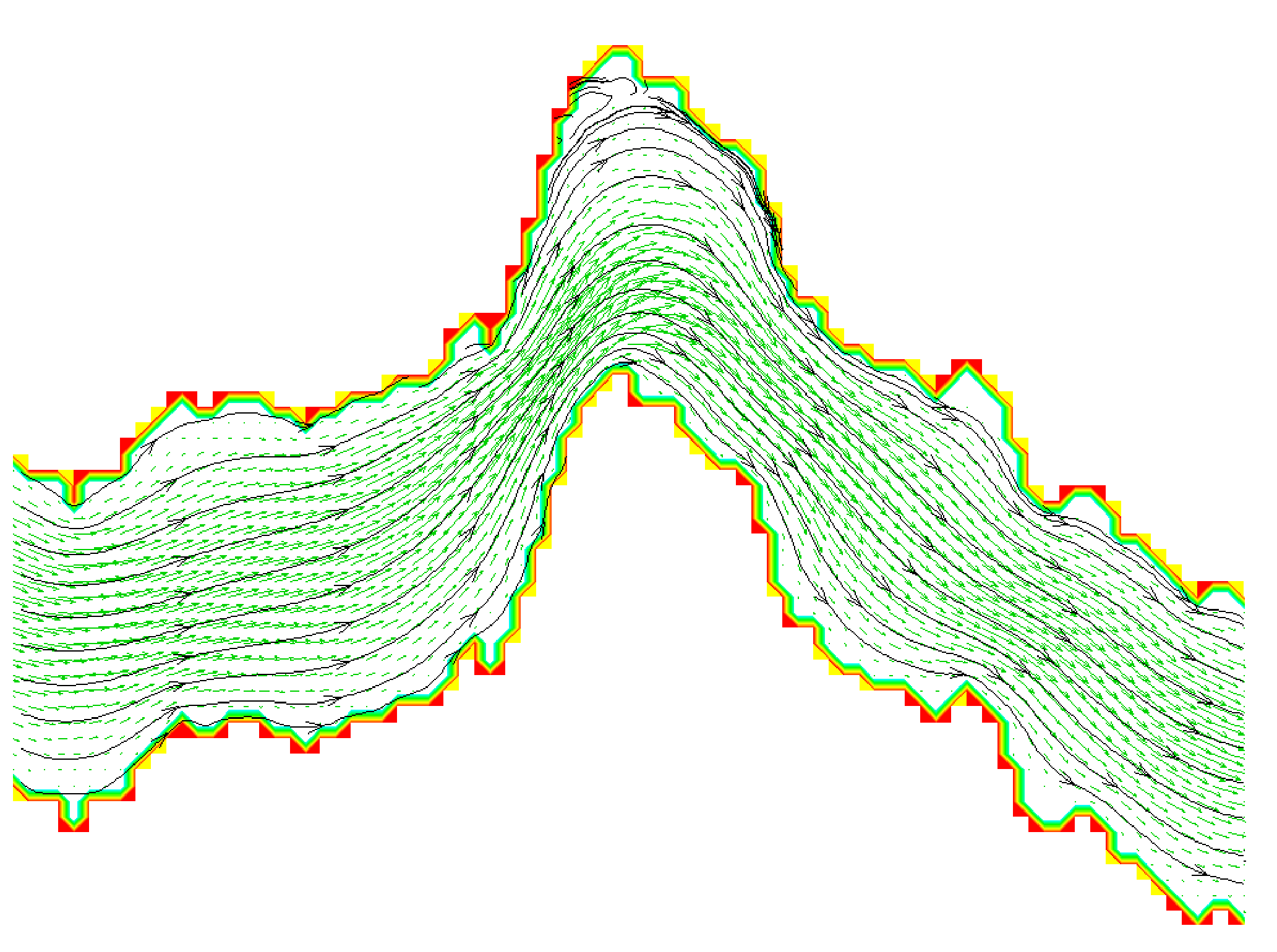}
\includegraphics[scale=.15]{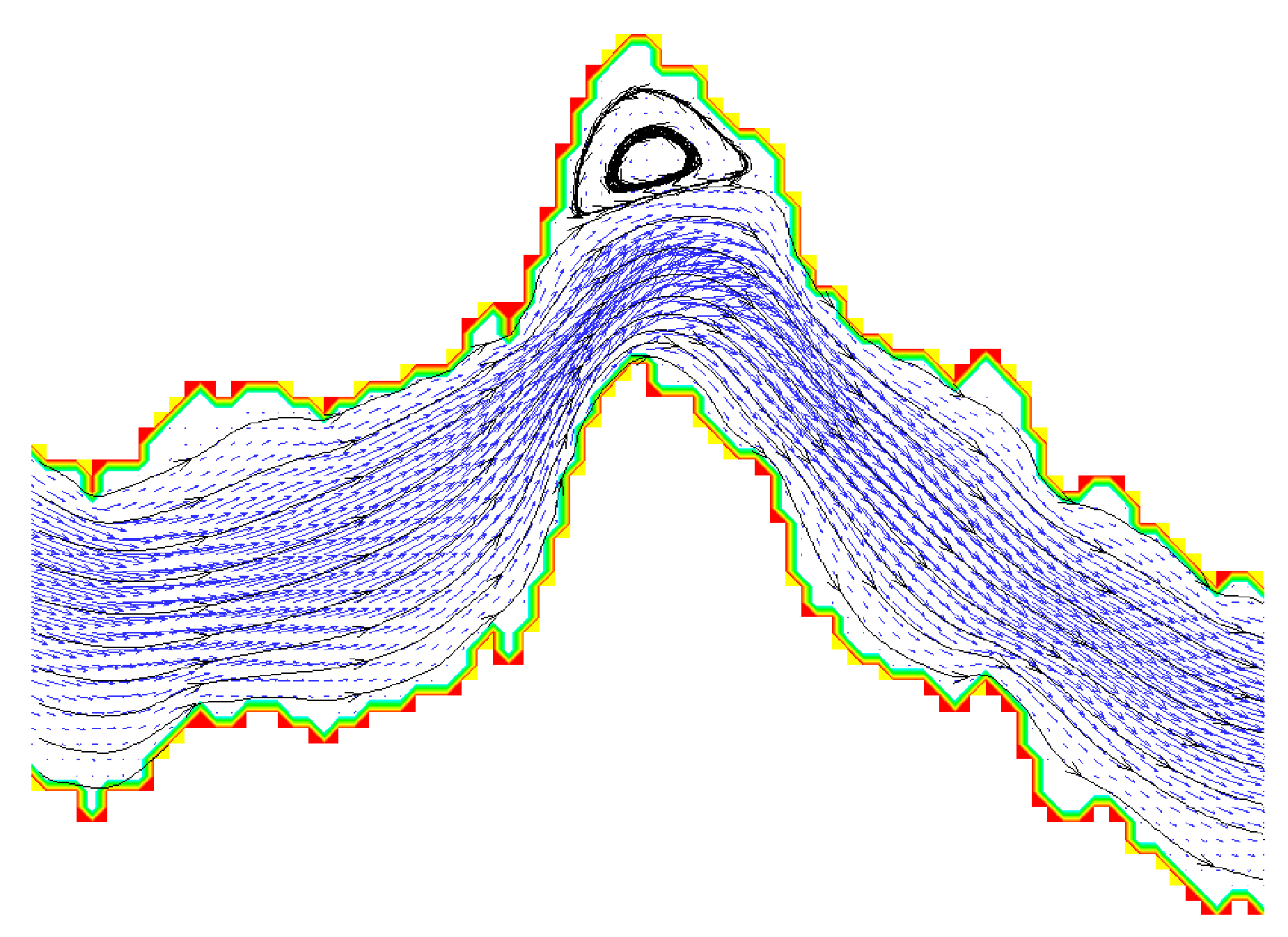}
\caption{Segment of Velocity vector field with streamlines of the flow for
power-law fluid with $m=0.01, n=0.75(top),1.0(middle),1.25(bottom)$ and the
pressure gradient applied is $G=1\times 10^{-6}$. The segment extends from
$x=20$ to $x=100$.}
\label{fig:vstream-75}
\end{center}
\end{figure}

We wish to examine how the local flow behavior varies with the
rheology of the fluid, at different geometrical features of a
self-affine channel.  We focus on a single realization of the
fracture, shown in Fig.~\ref{fig:saff-geom}, and vary the power law
index $n$ and the pressure gradient $G$.  The complete simulation
box has length $L=256$ in the flow direction and width $W=80$, in
terms of the (unit) lattice spacing, and the (constant) vertical
aperture is $H=20$. A uniform pressure gradient is
applied everywhere along the channel, as above, and periodic
boundadry conditions are applied in the flow direction. Local
minimum in the effective width (normal to the average flow) occur
around $x$=55, 110, and 240 where mass conservation implies the velocity magnitude
will be a maximum, irrespective of the rheology of the fluid.  In
Fig.~\ref{fig:vstream-75}, we show velocity fields and streamlines
for the three fluids along a segment of the fracture channel
$20\lesssim x\lesssim 100$ in Fig.~\ref{fig:saff-geom} which
includes a constriction, for applied pressure gradient $G=1\times
10^{-6}$. As we see, the streamlines are tortuous and very roughly
follow the channel walls, although recirculating eddies (closed
vortices) {\em may} occur where the channel exhibits side branches
or dead-end regions.  Indeed, at the present flow rate an eddy
appears in the shear-thickening case but not the others, presumably
because the velocites are higher in that case.

\subsection{Velocity field}

First we examine the variation of maximum absolute velocity along the channel, in
order to show how the fluid rheology influences the earlier results of Skjetne {\em et
al}. \cite{skjetne1999:flow} for the Newtonian case.  More precisely, for
each $x$ along the channel we compute the maximum over $y$ of $|\mathbf{u}(x,y)|$,
although we would have reached the same qualitative conclusions had we considered the
maximum over $y$ of $u_x(x,y)$.  Calculations were performed for three values
of the pressure gradient, $G=1\times 10^{-6}$, $5\times 10^{-5}$ and
$2\times 10^{-4}$, which correspond to Reynolds numbers $Re=0.95$, 37.0 and 92.7,
respectively, for the
Newtonian fluid.  Since the viscosity varies within the channel for the shear thinning
and thickening fluids, there is no unique definition of $Re$ in those cases, although
a convenient choice will be introduced in Section~\ref{sec:perm} for scaling purposes.

\begin{figure}[htb]
\begin{center}

\includegraphics[scale=.2,angle=-90.]{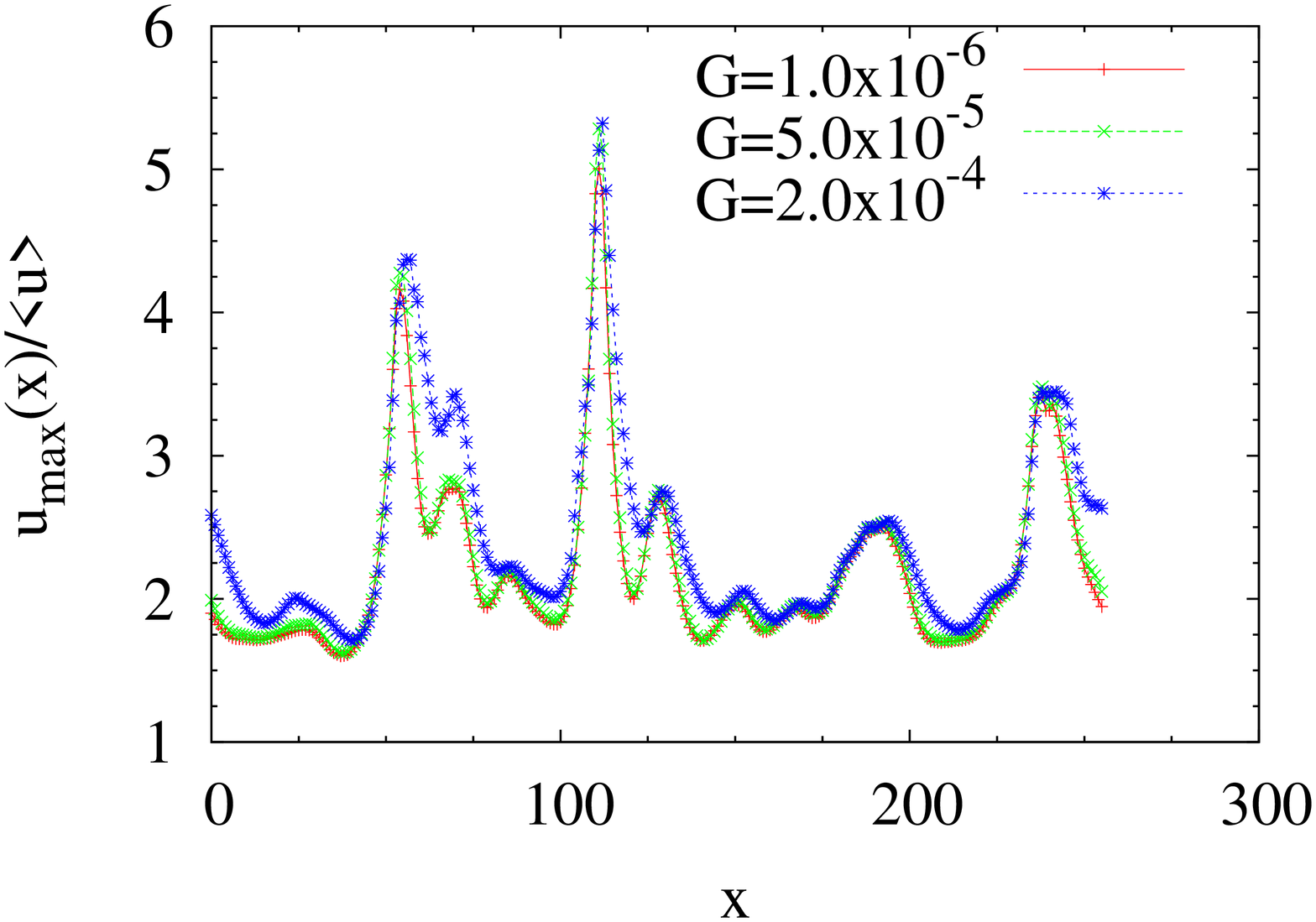}
\includegraphics[scale=.2,angle=-90.]{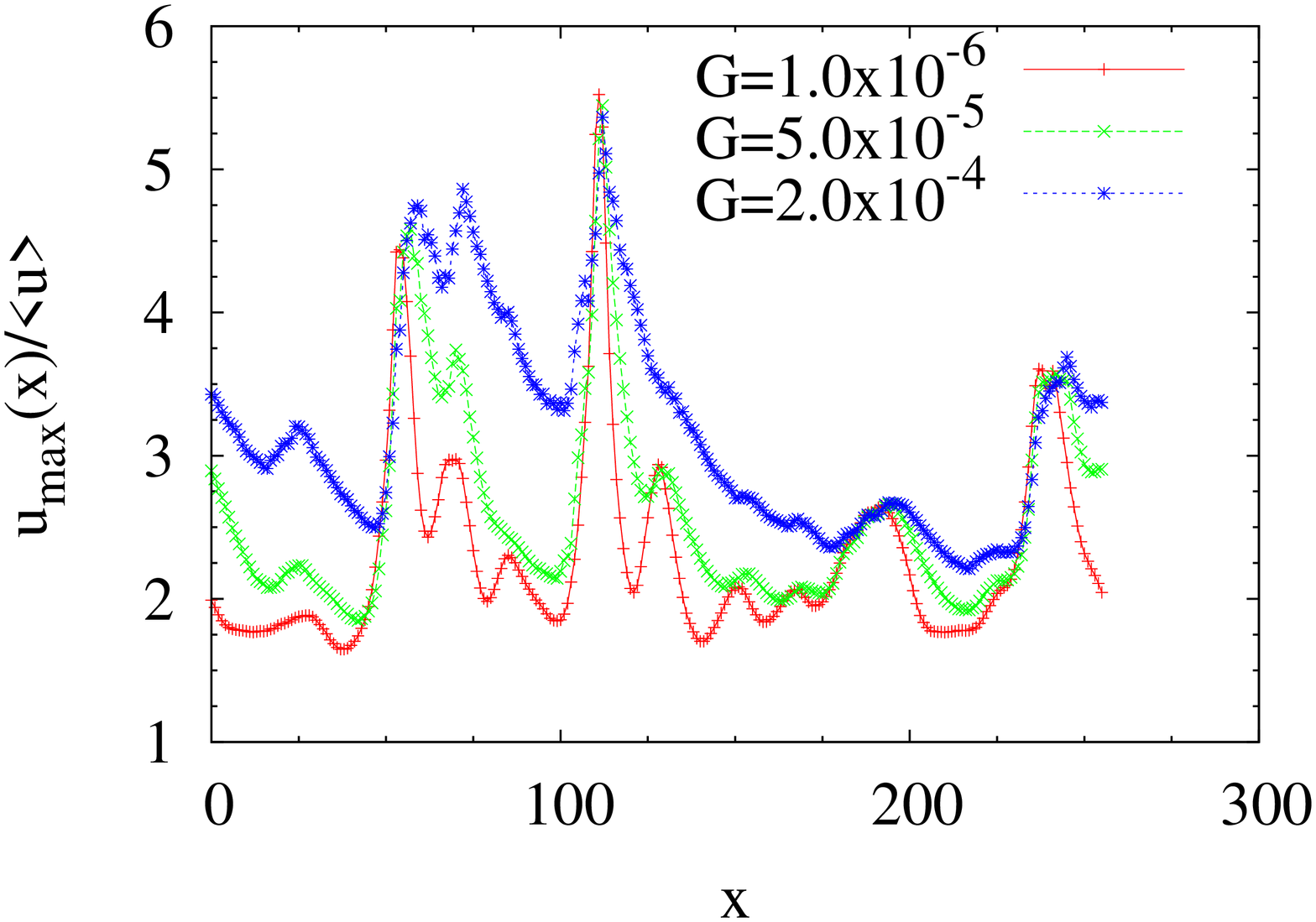}
\includegraphics[scale=.2,angle=-90.]{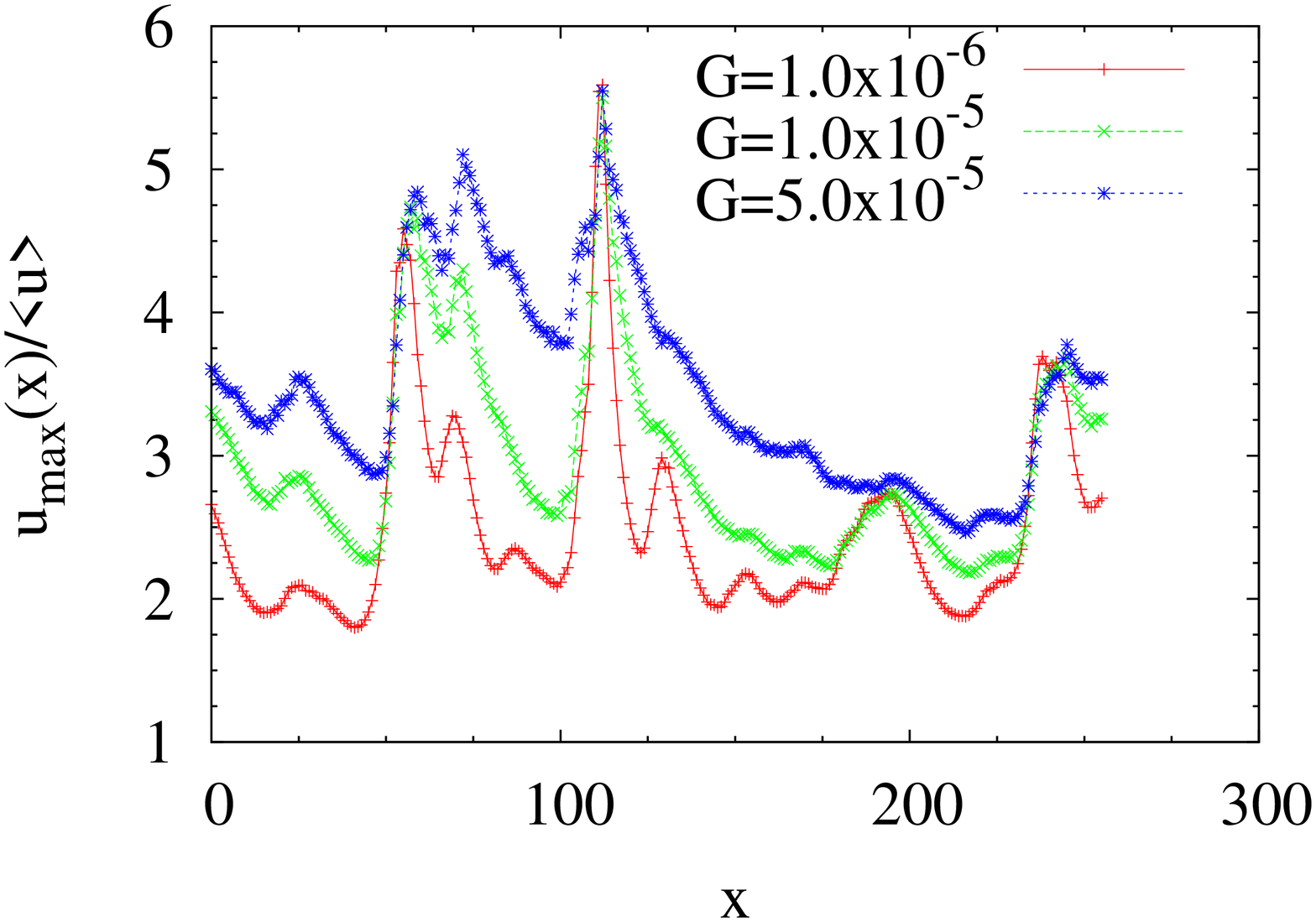}
\caption{Maximum absolute velocity along the fracture channel for shear-
thinning({\em top}, $n$=0.75),Newtonian({\em middle}, $n=1.0$), and shear-
thickening({\em bottom}, $n$=1.25) fluids for various applied pressure
gradient $G$. Each maximum velocity curve is normalized by the corresponding
$\overline{u_x}$,
the average flow velocity in the $x$-direction.}
\label{fig:maxv-n}
\end{center}
\end{figure}

The resulting plots of maximum velocity are shown in Fig.~\ref{fig:maxv-n}, where each
velocity is normalized by the average streamwise flow velocity $\overline{u_x}$
(referred to as
the interstitial velocity $u^*$ in \cite{skjetne1999:flow}), which equals the
flux divided by the channel width.
Obvious peaks appear at the positions of the visible constrictions in the channel near
$x=55$, 110 and 240, reflecting the narrowed aperture there. The normalized
peak heights are
fairly insensitive to the Reynolds number, although away from the peaks the trend is
for maximum velocity to increases with $Re$.  Note that for a flat  channel, the
the normalized maximum absolute velocity would equal 1.5, so the values of 5 or more
seen here are a substantial enhancement.  The peaks are not {\em all} closely correlated
with channel constrictions, however:  near $x=70$ and 130 maximum velocity peaks occur,
but at these locations the channel is expanding just downstram of a constriction.
It is also possible to calculate a ``maximum velocity trajectory'', following
\cite{skjetne1999:flow}, as the set of $(x,y)$ gridpoints which at each $x$ has the 
$y$-value corresponding to the position where the maximum velocity occurs.  For
the most part our observations concerning the behavior of these trajectories is
similar to that reported in this reference, but
we do not observe the line-length of this trajectory decreasing
monotonically with $Re$.

Comparing the other fluids to the Newtonian case, we see in Fig.~\ref{fig:maxv-n} that
the global maximum absolute velocity always occurs at the narrowest constriction near
$x=110$  and the other primary peaks always occur at the same positions, $x=55$ and
240, as well.  Furthermore, each peak has roughly the same (normalized) velocity
value.  In the shear-thinning case, both the variation in $x$ away from the
peaks/constrictions and the variation with pressure gradient are weaker than in the
other cases, which may be attributed to the fact that typical velocities in the
fracture are smaller in this case, and inertial effects play a weaker role.
In the shear-thickening case, where typical
velocites are larger, the maximum velocity values are larger off the peaks values,
and furthermore exhibits rather more variation with $x$ and $Re$ than the other fluids.

\begin{figure}[htb]
\begin{center}
\includegraphics[scale=.2,angle=-90.]{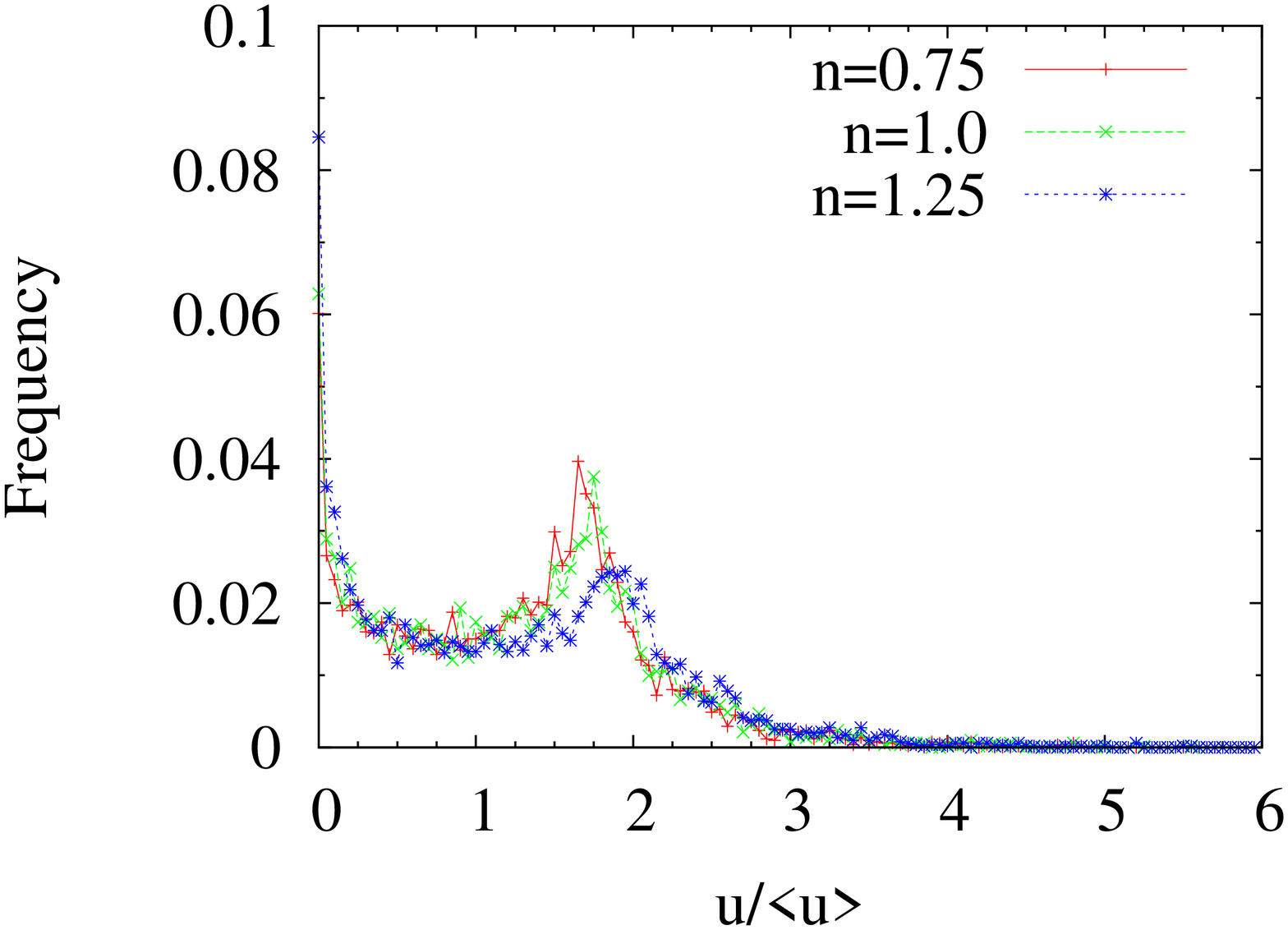}
\includegraphics[scale=.2,angle=-90.]{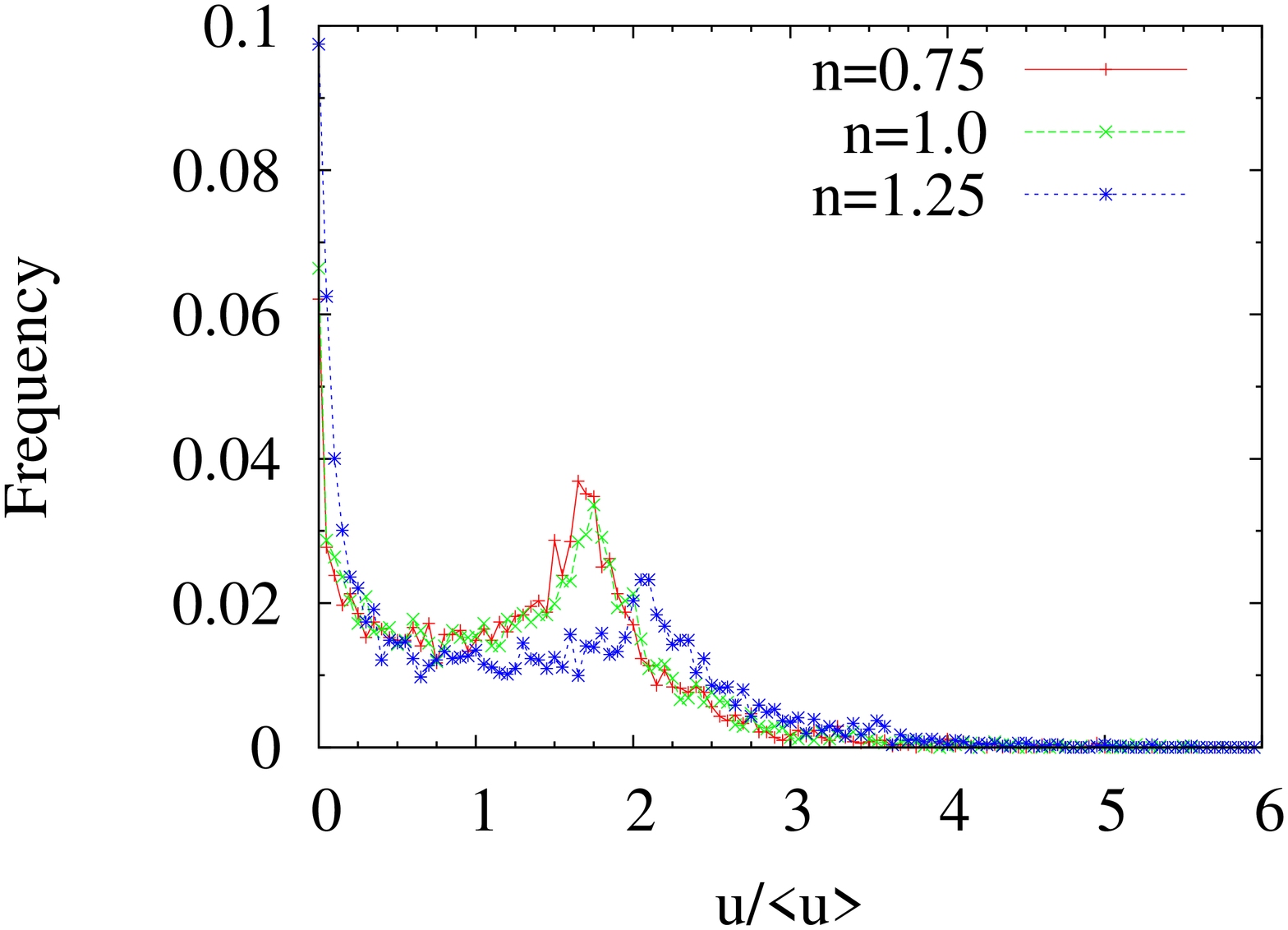}
\includegraphics[scale=.2,angle=-90.]{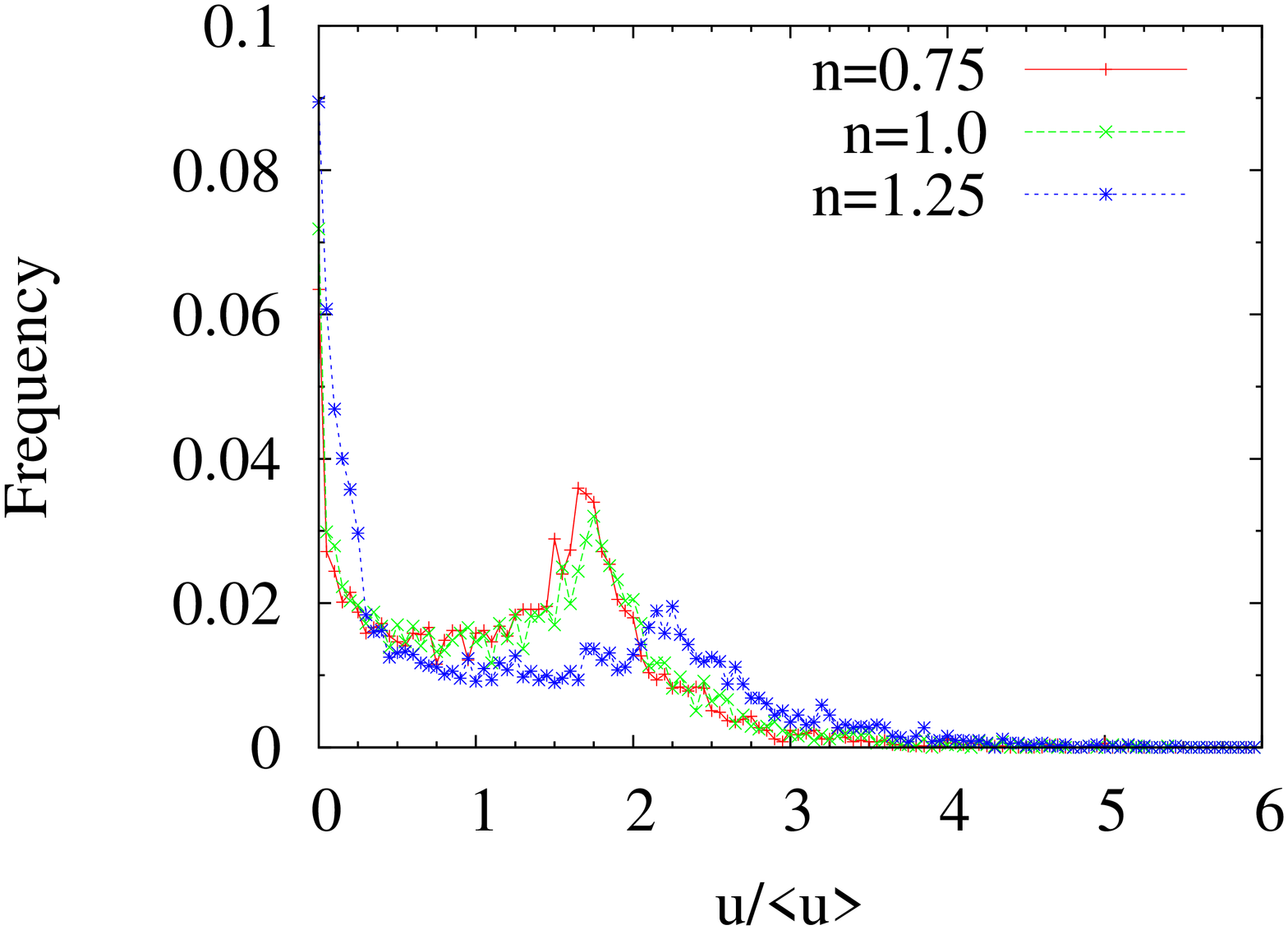}
\caption{Distribution of normalized absolute velocity in the whole self-affine
fracture flow domain for different power-law fluids with pressure gradients
$G=1.0e-6$, $5.0e-6$ and $1.0e-5$ (top to bottom).}
\label{fig:velfreq}
\end{center}
\end{figure}

The probablity distribution of velocity magnitudes is also of interest
\cite{skjetne1999:flow},
since the presence of low and high velocity components strongly influences mixing
processes and transport of passive tracers and suspended particles \cite{guyon}.
Histograms of the observed absolute value of the velocity for the three fluids at
various pressure gradients are shown in Fig.~\ref{fig:velfreq}.  In all cases there is
a peak near the origin, which reflects the numerous low-velocity zones in the
crevasses at the fractures walls, along with a higher-velocity peak resulting from the
rapid flow in the channel constrictions.  The latter moves out to higher values as the
pressure gradient increases (note the normalization by $\overline{u_x}$ in the figure)
Once again, the shear-thickening case behaves somewhat differently than the other two
fluids, showing a less prominent and broader ``constriction peak,'' and more variation
with $G$.

\subsection{Pressure and stress field}
\begin{figure}[htb]
\begin{center}
\includegraphics[scale=.3,angle=-90.]{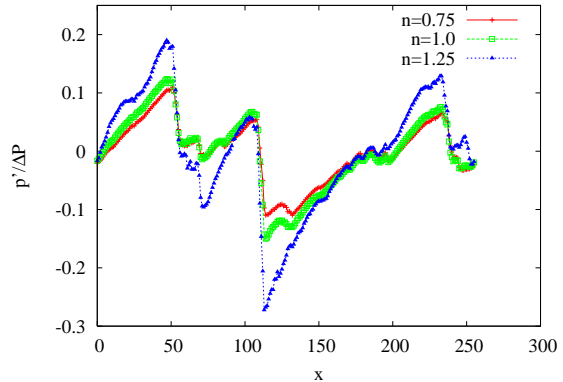}
\caption{Normalized pressure fluctuation along the channel for different power-law
fluids, at pressure gradient $G=1.0e-5$.}
\label{fig:p-diff-n}
\end{center}
\end{figure}

The distribution of pressure and stress in the fluid are important for non-Newtonian
rheology, and in considering possible erosive processes on the fracture walls.
To contrast the behavior of the different fluids, Fig.~\ref{fig:p-diff-n} shows the
pressure ``fluctuations'' along the channel for the three power-law fluids $n=0.75$,
1.0 and 1.25. The fluctuation $p'$ is the deviation in pressure from the imposed
linear gradient, which would vanish identically in a Hele Shaw geometry. In the
figure, the fluctuation has been normalized by the imposed pressure difference,
$\Delta p = GL$, and averaged over the channel width.
For all three fluids, the pressure fluctuation are most significant in the vicinity of
the main constrictions in the channel where the fluid accelerates, rising just before
each constriction's location
and dropping rapidly as it is traversed.  Some additional structure arises at
positions $x$ of bends in the flow path, another source of fluid acceleration.
Again, the shear-thinning and Newtonian
fluids behave somewhat similarly, while the variation is strongest in the
shear-thickening case.

\begin{figure}[htb]
\begin{center}
\includegraphics[scale=.2,angle=-90.]{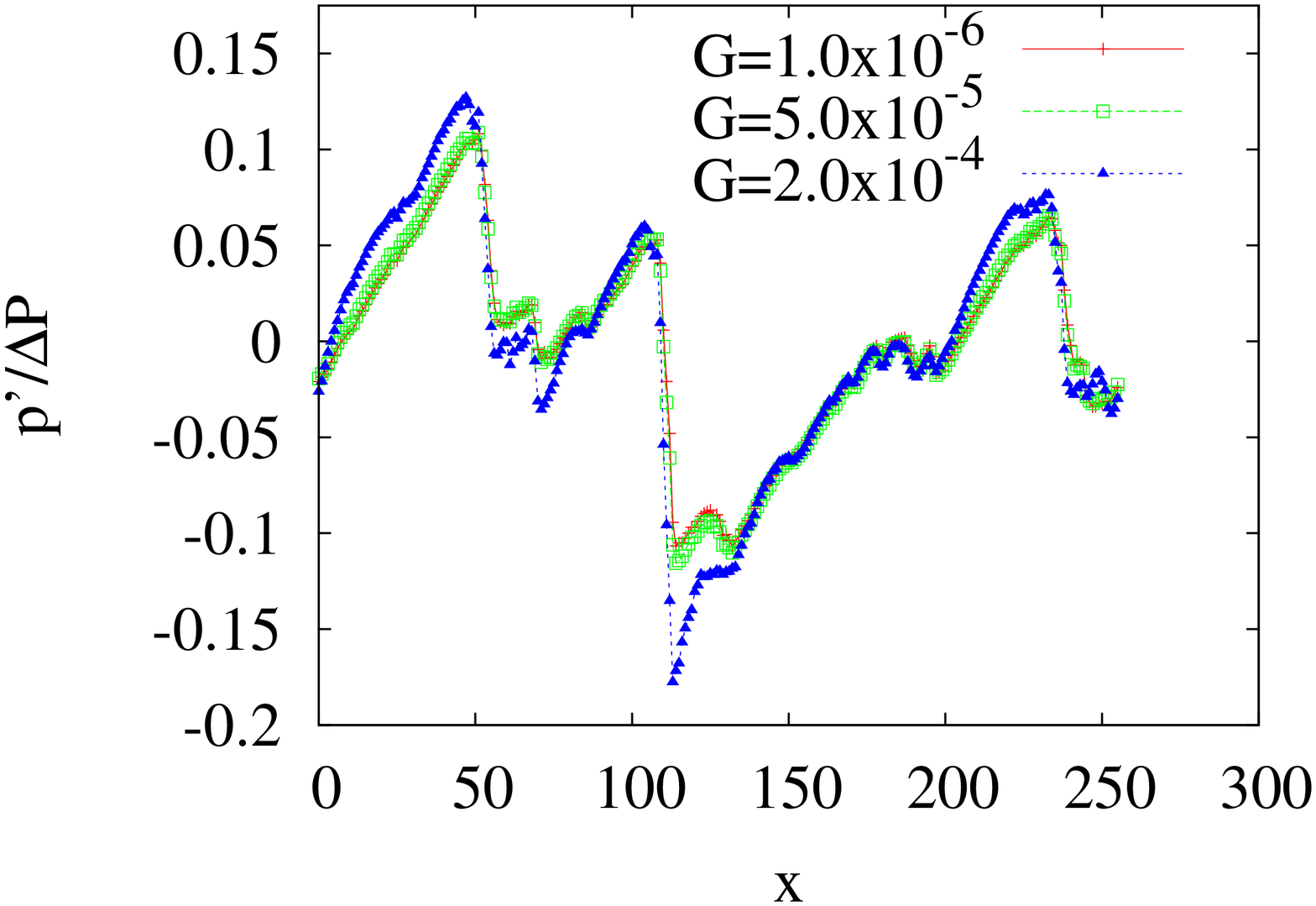}
\includegraphics[scale=.2,angle=-90.]{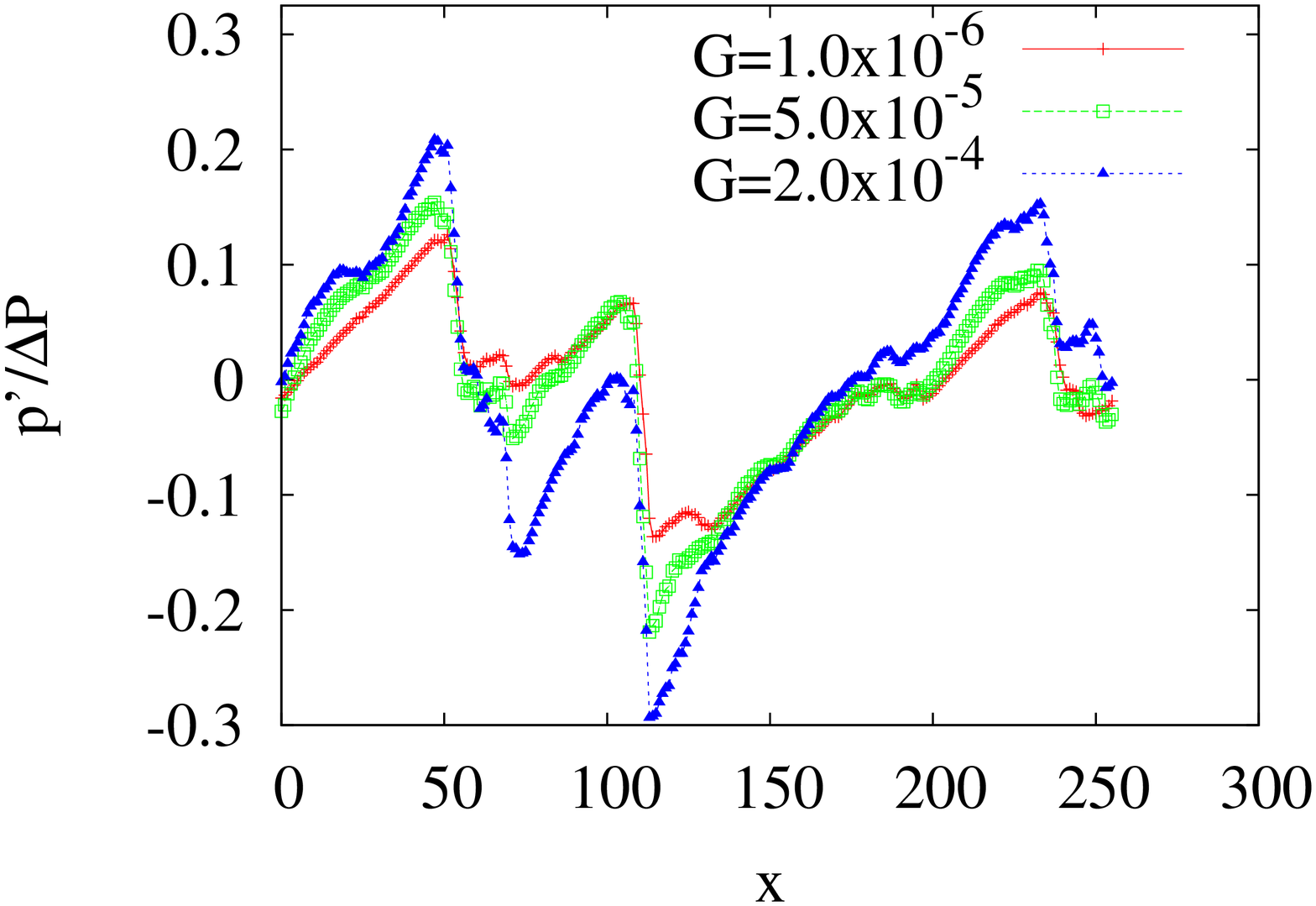}
\includegraphics[scale=.2,angle=-90.]{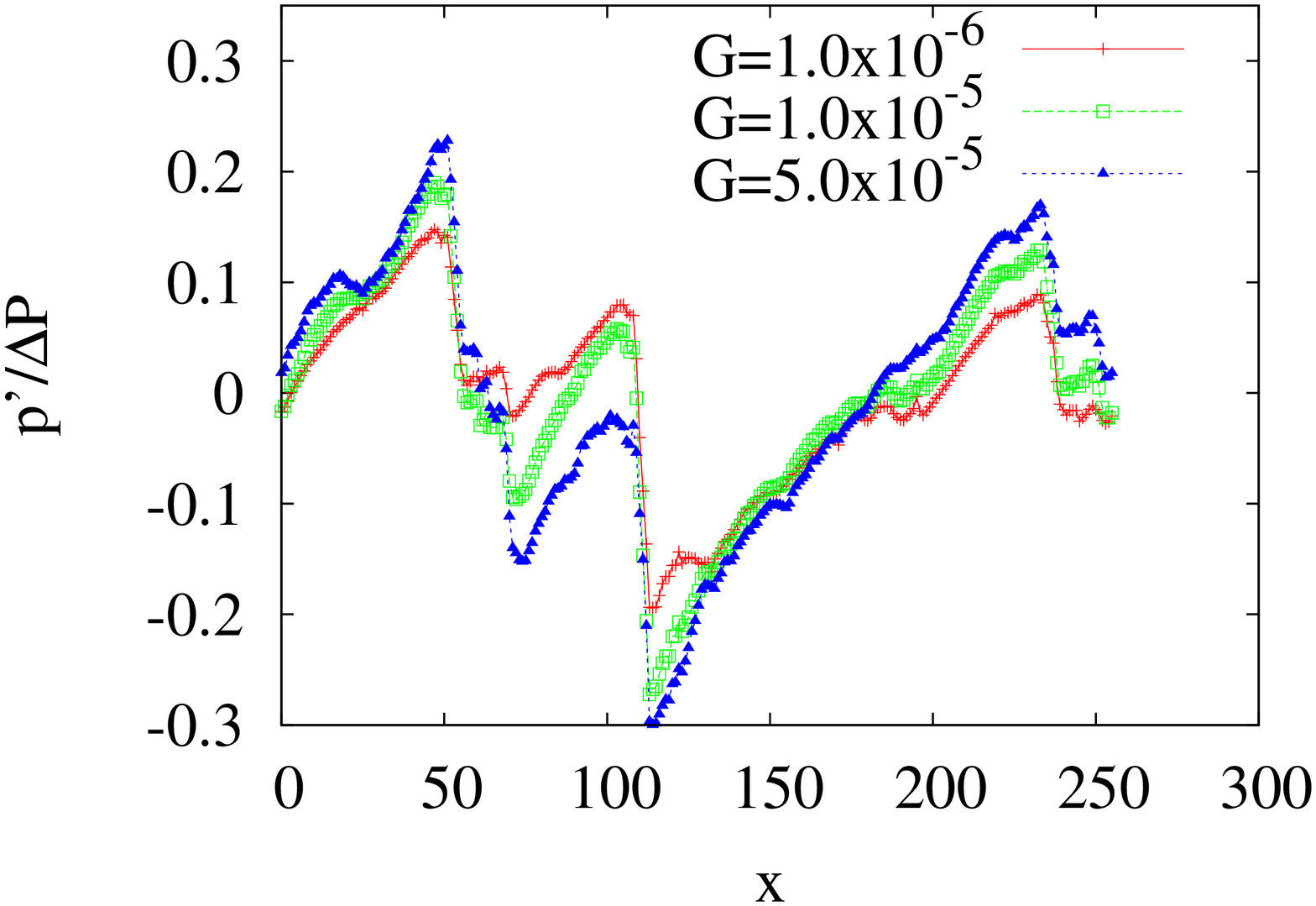}
\caption{Pressure fluctuations of different fluids with power $n=0.75, 1.0,
1.25$ and $m=0.01$ along a self-affine fracture channel under different applied
pressure gradients.}
\label{fig:p-n}
\end{center}
\end{figure}

The variation in fluctuation with imposed gradient is shown in
Fig.~\ref{fig:p-n}, and indicates the expected general increase in magnitude with $G$
along the channel.

To assess the effects of the flow on the fracture wall, we first calculate the average
force exerted by the fluid on the wall,
\begin{equation}
\mathbf{F} = {1\over L}\int d\ell\, \mbox{\boldmath $\hat{n}\cdot\sigma$}
\end{equation}
where the integral runs over the fracture surface (a curve in this two-dimensional
calculation), and $\mathbf{\hat{n}}$ is the local normal to the wall. The force is
then decomposed into $x$ and $y$ components, representing the average drag and thrust on
the wall, respectively, and then normalized by a typical inertial pressure $\rho
\overline{u}^2/2$ times the nominal surface area $L\times 1$, to give drag and thrust
coefficients
\begin{equation}
 d=\frac{F_x}{L\rho \overline{u}^2/2}, \qquad t=\frac{F_y}{L\rho \overline{u}^2/2}.
\end{equation}
Note that aside from the (reasonable) use of the inertial pressure, the remainder of the
normalization is somewhat arbitrary but a fixed constant for each fracture, and
mainly serves to provide dimensionless drag and thrust coefficients.
The drag and thrust forces for the lower and upper walls of the channel are similar
but not identical because of the asymmetry of the fluid-solid boundary, and for
definiteness we present only the forces on the lower wall.

\begin{figure}[htb]
\begin{center}
\includegraphics[scale=.3,angle=-90.]{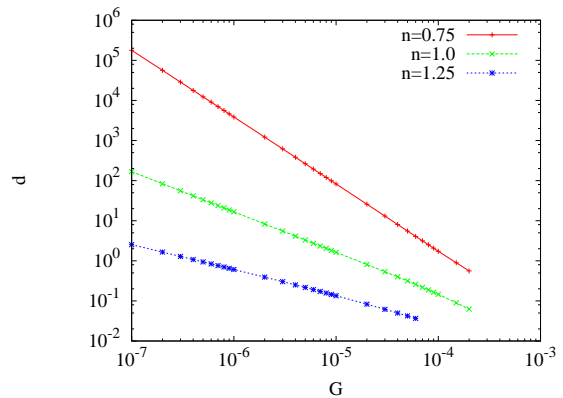}
\caption{Drag factor $d$ for power-law fluids in a self-affine fracture channel
as a function of applied pressure gradient.}
\label{fig:drag}
\end{center}
\end{figure}

\begin{figure}[htb]
\begin{center}
\includegraphics[scale=.3,angle=-90.]{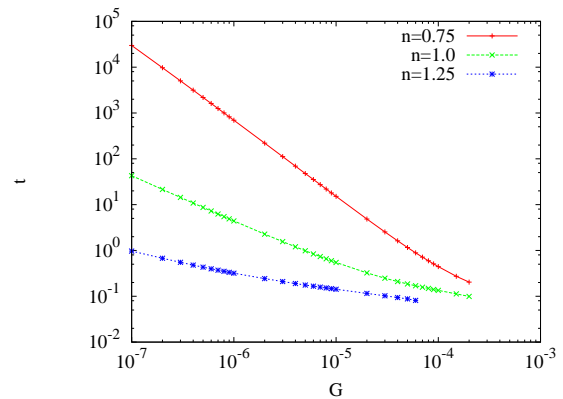}
\caption{Thrust factor $t$ for power-law fluids in a self-affine fracture channel
as a function of applied pressure gradient.}
\label{fig:thrust}
\end{center}
\end{figure}

The results of calculating the drag and thrust coefficients is shown in
Figs.~\ref{fig:drag} and \ref{fig:thrust} for the three fluids with exponents
$n=0.75,1.0,1.25$.  In all cases, the coefficients exhibit simple power-law behavior,
provided $G$ is not too large,
and the transition to a different behavior at larger $G$ may be associated with the
onset of inertial effects (see the following section).
This form of scaling behavior result is consistent with the experimental results
reported in
\cite{chhabra2001:review}, and the values of the slopes found in the log-log plots in
the low-$G$ range,
-1.67, -1.02, -0.62 for $n=0.75$, 1.0 and 1.25, respectively, for {\em both} drag and
thrust, may be understood from the following argument.

If inertial effects are absent, one expects the scaling behavior in a rough channel
to be the same as in a straight channel.  In that case, from Eq. (\ref{eqn:vpower-law})
one has $u\sim G^{1/n}$, and therefore $\nabla u \sim G^{1/n}$
as well, so that $\mu\sim|\nabla u|^{n-1}\sim G^{(n-1)/n}$.  The drag and thrust
forces are
proportional to the stress, $\sigma\sim \mu\nabla u\sim G^{(n-1)/n+1/n}\sim G^1$.
The drag and thrust {\em coefficients} are then
$d,t\sim F_{x,y}/\overline{u}^2\sim \sigma/u^2\sim G^{1-2/n}$,
giving exponents -5/3, -1 and -3/5, respectively, for the three fluids.


\section{Permeability}
\label{sec:perm}
Next we consider global behavior -- the permeability of a self-affine fracture channel.
Our discussion is colored by analogies to flow in intergranular porous media,
so we first recall the situation in that system \cite{bear}.
For Newtonian fluids in the low Reynolds number limit, the definition of intergranular
permeability is given by Darcy's law, $\langle\mathbf{u}\rangle = -(k/\mu) \nabla p$, where
$\langle\mathbf{u}\rangle$ is the average flow velocity and $p$ the average pressure. 
The average
in question could be a volume average or an ensemble average, and for a flow which is
macrosopically unidirectional, an operational definition of permeability is
$k=\mu Q L/A\Delta p$ where $Q$ is the flux through a sample of cross-sectional area
$A$ and length $L$.  In a two-dimensional situation, the area is replaced by the width
$W$, and $Q$ is the flow per unit length in the third direction.  A definition identical
to the latter case may be used for the permeability of low Reynolds number Newtonian
flow in a fracture.  Both finite Reynolds number flow and non-Newtonian fluid rheology
modify this description.  We first consider the effects of inertia, and then examine
how permability relates to the fracture morphology.

\subsection{Inertial effects}

At higher flow rates when inertial effects appear, the
relation between pressure difference and average velocity or flux becomes nonlinear
and one may write
\begin{equation}
\Delta p = \alpha Q + [\,\beta Q^2\mbox{  or }\gamma Q^3]
\label{eq:forch}
\end{equation}
where $\alpha$ incorporates the Darcy permeability, and the term in brackets is
the inertial correction, with $\alpha,\beta >0$.  At high $Q$ the quadratic
or ``Forchheimer'' term applies,
but in the transitional region where the Reynolds number is small but finite, a
cubic dependence is
found. This picture is supported by experiments, analytic calculations, and numerical
simulations \cite{koch}.

\begin{figure}[htb]
\begin{center}
\includegraphics[scale=.3,angle=-90.]{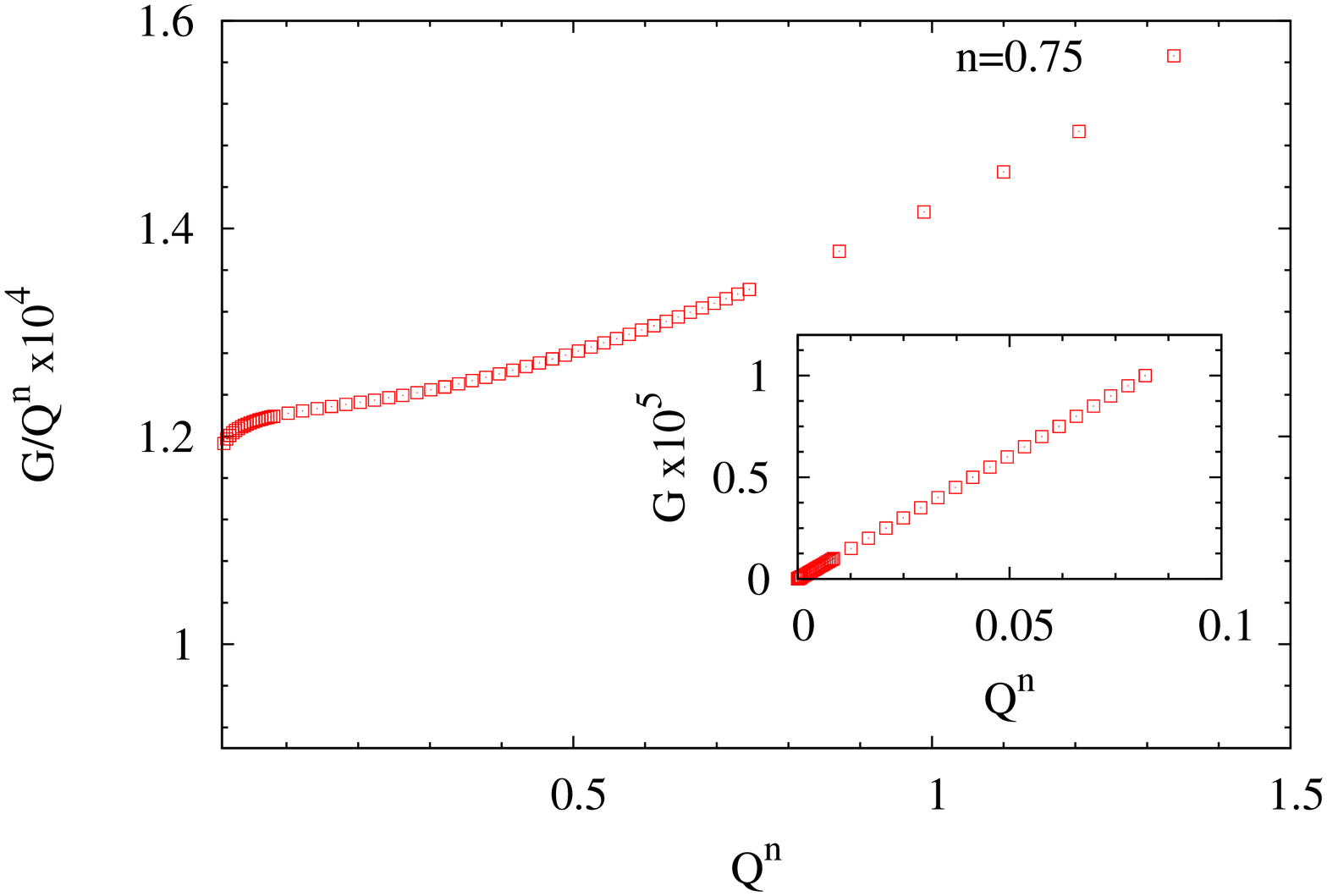}
\includegraphics[scale=.3,angle=-90.]{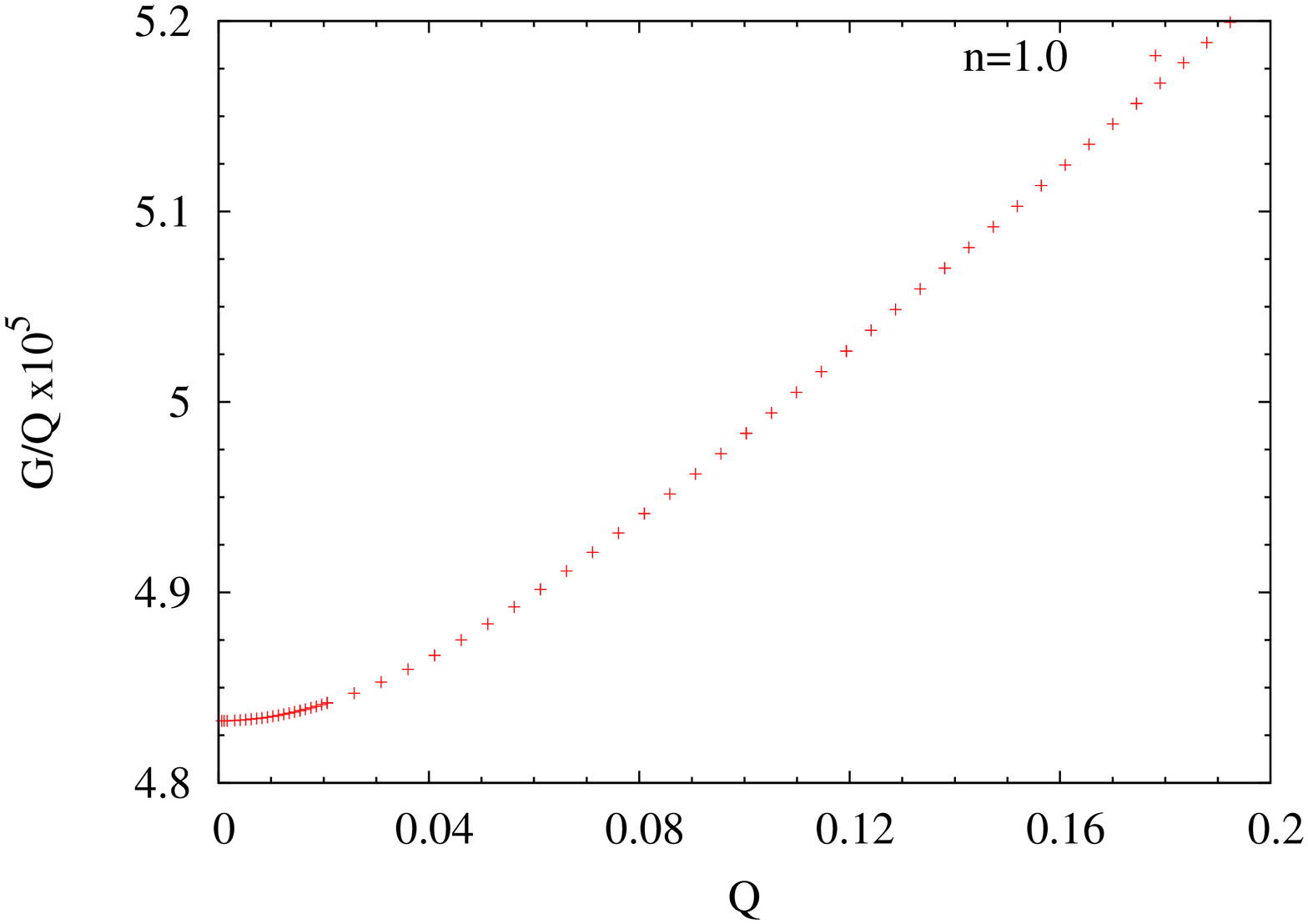}
\includegraphics[scale=.3,angle=-90.]{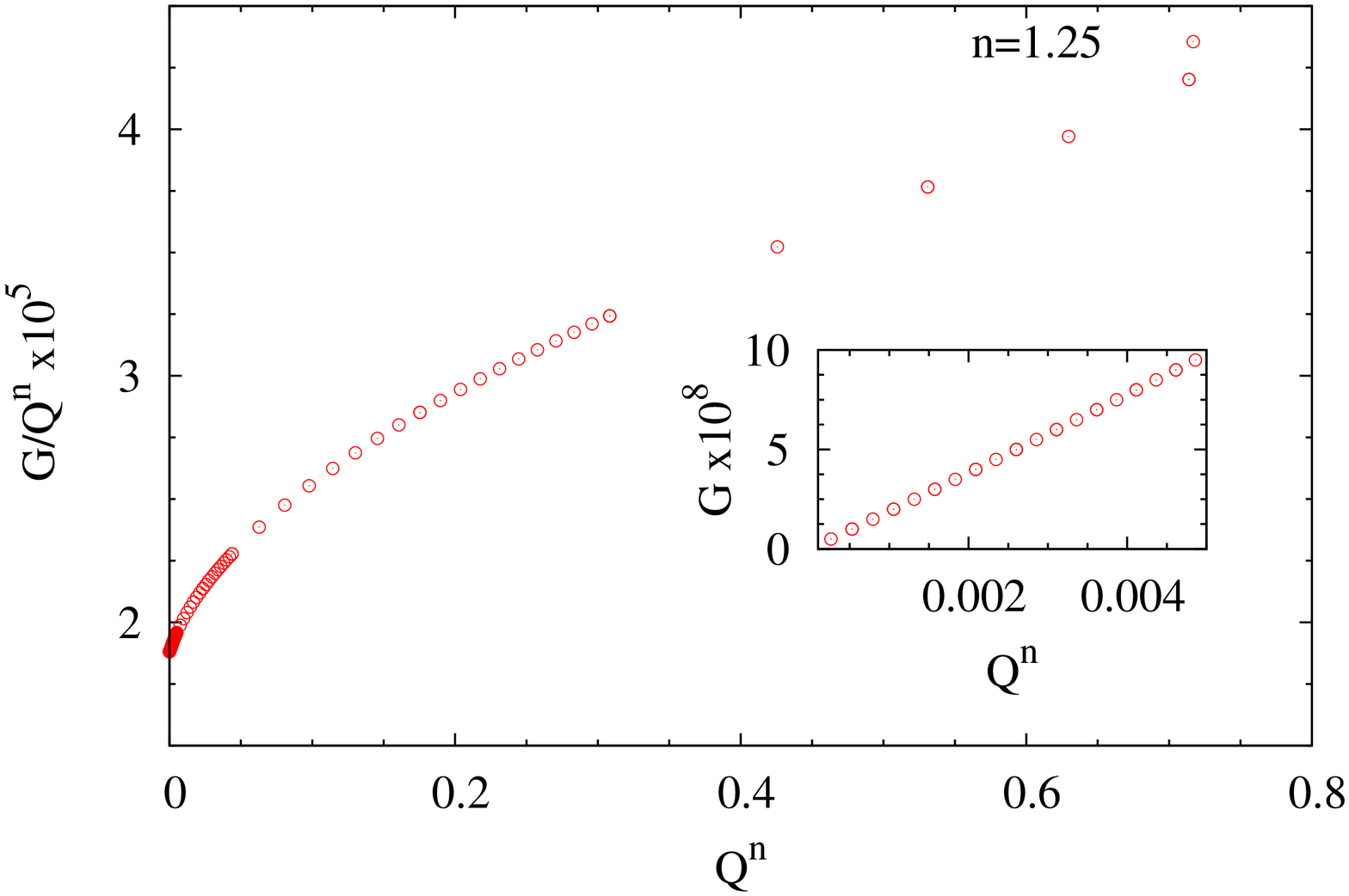}
\caption{Relation between imposed pressure gradient and fluid flux for power law
fluids:  $n=0.75$, 1.0 and 1.25 (top to bottom).
\label{fig:trans-1}}
\end{center}
\end{figure}

The flow of a Newtonian fluid in a self-affine fracture can be described in identical
terms, as
shown by the numerical simulations of Skjetne {\em et al}. \cite{skjetne1999:flow}
which exhibit the same transitions between flow regimes indicated in
Eq.~(\ref{eq:forch}).
In extending the discussion to power-law fluids, the first issue is to choose the
appropriate power of $Q$.
The exact solutions for Hele Shaw flow given in Eq.~(\ref{eq:flux}) have the scaling
behavior $G\sim Q^n$ where $G$ is the applied pressure gradient (the relevant pressure
for macroscopic behavior) and $n$ the power-law index.  In a rough fracture, one would
naturally expect an identical relation, albeit with a modified coefficient, at
low $G$, and then at larger $G$ inertial effects would be expected to produce
(positive) terms involving higher powers of $Q$.  To test this idea, note that
we are concerned here with the statistical
behavior of self-affine fractures, rather than the details of flow in one particular
geometry which was relevant in the previous section, so
an ensemble average over six realizations of the fracture surface is used.
The simulation results are shown in Fig.~\ref{fig:trans-1}
and indeed show a $G\sim Q^n$ scaling behavior at low $G$.  The Newtonian $n$=1 plot
shows this behavior clearly since $G/Q^n$ is constant at low $Q$, whereas in the other
cases, the expected behavior is present at sufficently small $Q$ as indicated in the
alternative plots in the insets of $G$ {\em vs}. $Q$.
The need for different plotting variables arises becuase in the non-Newtonian cases,
the flow rate fluctuates substantially at low $G$ and division by $Q^n$ is numerically
unstable.  Beyond the quasi-linear regime, the Newtonian case shows the expected
trasition to a Forchheimer flow regime $G\sim Q^2$ at larger forcing, and the
shear-tickening fluid shows a somewhat analogous behavior $G\sim Q^{2n}$.  The
shear-thinning fluid is not described by a simple power law at large $G$, and we are
not aware of any theoretical treatment of this problem, so we simply report the
numerical results.

To understand the numerical coefficient in the flow results, the fracture-modified
Darcy permeability, we again refer to the Hele Shaw case and define
\begin{equation}
k=\frac{m^{1/n} \overline{u}}{G^{1/n}}.
    \label{eq:k-power-law}
\end{equation}
Since the roughness and tortuosity of the fracture cause the stramlines to bend and
viscous dissipation to increase, the permeability should be reduced compared to a
smooth and flat Hele Shaw geometry of the same aperture.  In Table I, the various
permeabilites are compared, and a reduction by a factor 6-7 is found.

\begin{table}
\begin{center}
\begin{tabular}{|l|c|c|} \hline
fluid index &   $k_0$   &   $k$ \\ \hline
$n=0.75$    &   2.99    &   0.373   \\ \hline
$n=1.0$     &   33.2    &   4.73    \\ \hline
$n=1.25$    &   142 &   20.6    \\ \hline
\end{tabular}
\end{center}
\caption{Effect of roughness and tortuosity on the low Reynolds number permeabilty:
$k_0$ and $k$ are the permeabilitites (defined in Eq.~(\ref{eq:k-power-law}) for a
Hele Shaw cell and a self-affine fracture of the same mean aperture, respectively.}
\end{table}

\begin{figure}[htb]
\begin{center}
\includegraphics[scale=.3,angle=-90.]{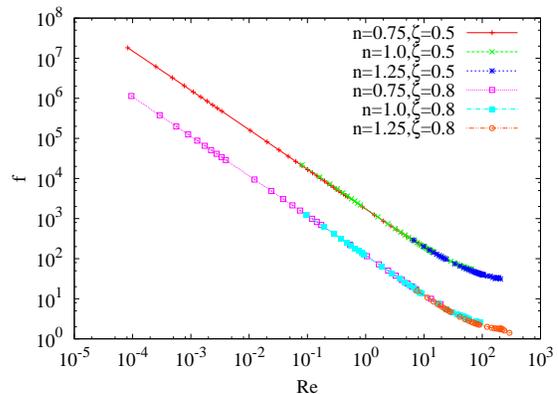}
\caption{Friction factor of self-affine fracture channels of Hurst exponent
$\zeta=0.5$ and $0.8$ as a function of Reynolds number defined as in \ref{eq:re}
for power-law fluids with $n=0.75,1.0,1.25$. \label{fig:re-f-zeta-n}}
\end{center}
\end{figure}

So far, we have expressed the pressure gradient $G$ in terms of the flux $Q$,
because these quantities are well defined in the present simulations.  However, for
general purposes It is preferable to use a dimensionless quantity such as the Reynolds
number as the independent variable, but the definition of $Re$ for power-law fluids is
not entirely obvious for power-law fluids because the
the viscosity varies over the flow domain.  One way to combine the results for
different fluids is based on an analogy to the friction factor scaling laws for flow
in pipes originally due to Nikaradze \cite{schlichting}, which can be extended to
non-Newtonian fluids as shown by Metzner \cite{metzner}.  Recall that
for unidirectional flow of a Newtonian fluid of viscosity $\mu$ in a pipe of diameter
$D$, the mean velocity is $\overline{u} = GD^2/32\mu$ and the shear stress at the wall is
$\tau_w = GD/4$, so if one defines the conventional friction factor as
$f=\tau_w/{\scriptstyle {1\over 2} }\rho\overline{u}^2$, then one finds $f=16/Re$
where $Re=\rho\overline{u}D/\mu$.
Experiments follow this scaling law up to a value of $Re$ that depends on the roughness
of the pipe, and at larger values of $Re$, $f$ levels off.  An analogous calculation
for Hele Shaw flow using the aperture $H$ instead of the diameter $D$ gives
$\tau_w=GH/2$ and
$f=12/Re$.  The power-law generalization is to use the latter form for $\tau_w$,
along with Eq.~(\ref{eq:flux}) to express the pressure gradient in terms of the mean
velocity, and yields
\begin{equation}
f={12\over Re}\qquad\mbox{if}\quad Re \equiv 6\rho \overline{u}^{2-n}H^n/m',
\label{eq:re}
\end{equation}
where $m'=m (2(2n+1)/n)^n$.  This choice of variables is not the last word,
because in the analogous interganular porous medium case where a similar approach
has been taken \cite{chhabra2001:review}, extra constant factors such as functions of
the porosity or the ``dynamic specific surface area'' are introduced into the friction
factor and Reynolds number definitions to promote data collapse.  It is not clear how
such ad hoc factors might be interjected here, so instead we collapse the data using
a simple constant factor which varies from fluid to fluid, and the result is shown in
Fig.~\ref{fig:re-f-zeta-n}.  Two different values of the Hurst exponent are shown, and
in both cases we see an $F\sim 1/Re$ scaling at low $Re$, a transition at $Re\sim 1-10$
and perhaps a constant friction factor at larger $Re$.  Unfortunately, the
calculations cannot be extended into the latter regime using the present
method (a particluar implementation of the lattice Bolzmann technique) becuase
numerical instabilites arise.

\subsection{Morphology effects}
\begin{figure}[htb]
\begin{center}
\includegraphics[scale=.3,angle=-90.]{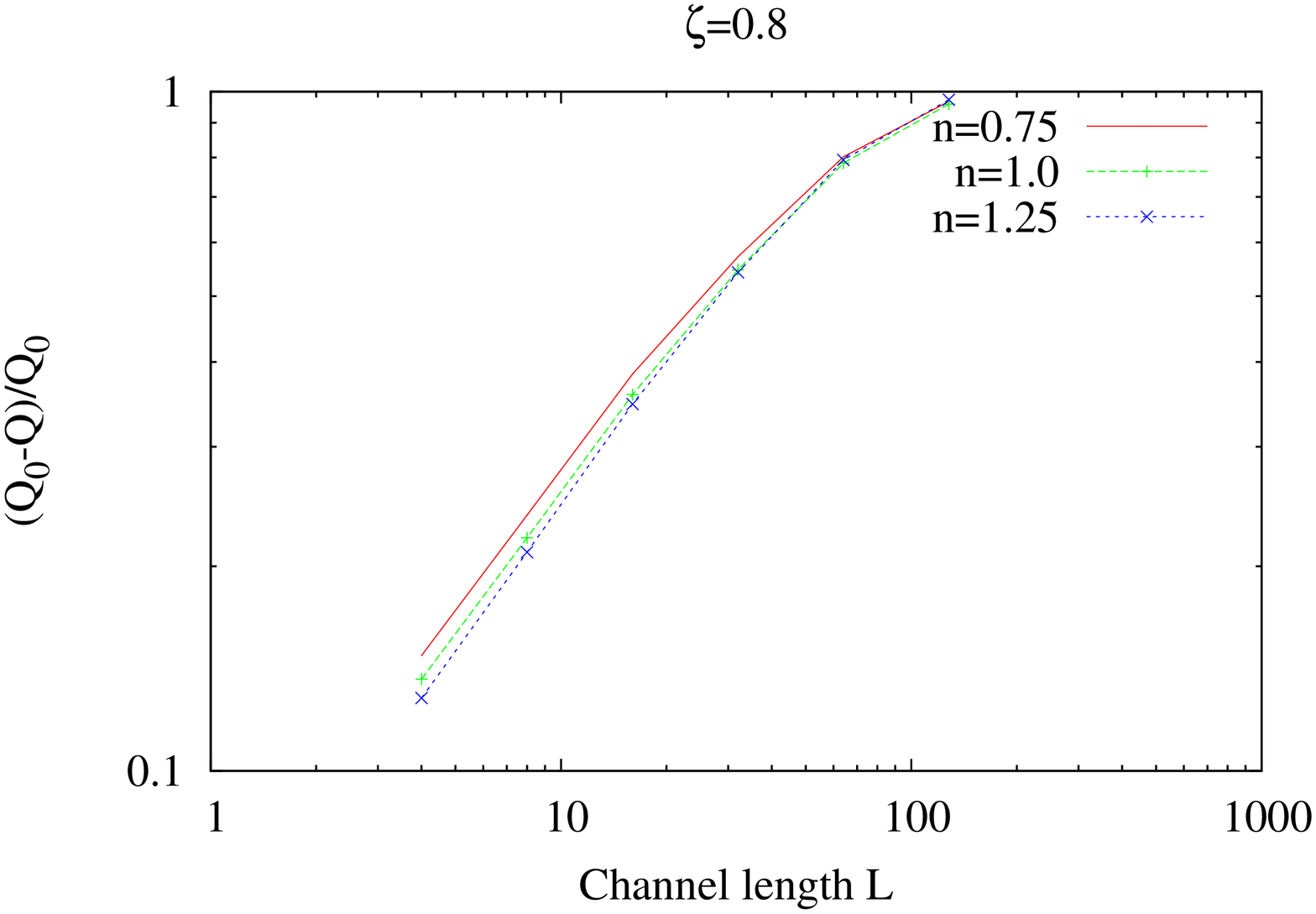}
\includegraphics[scale=.3,angle=-90.]{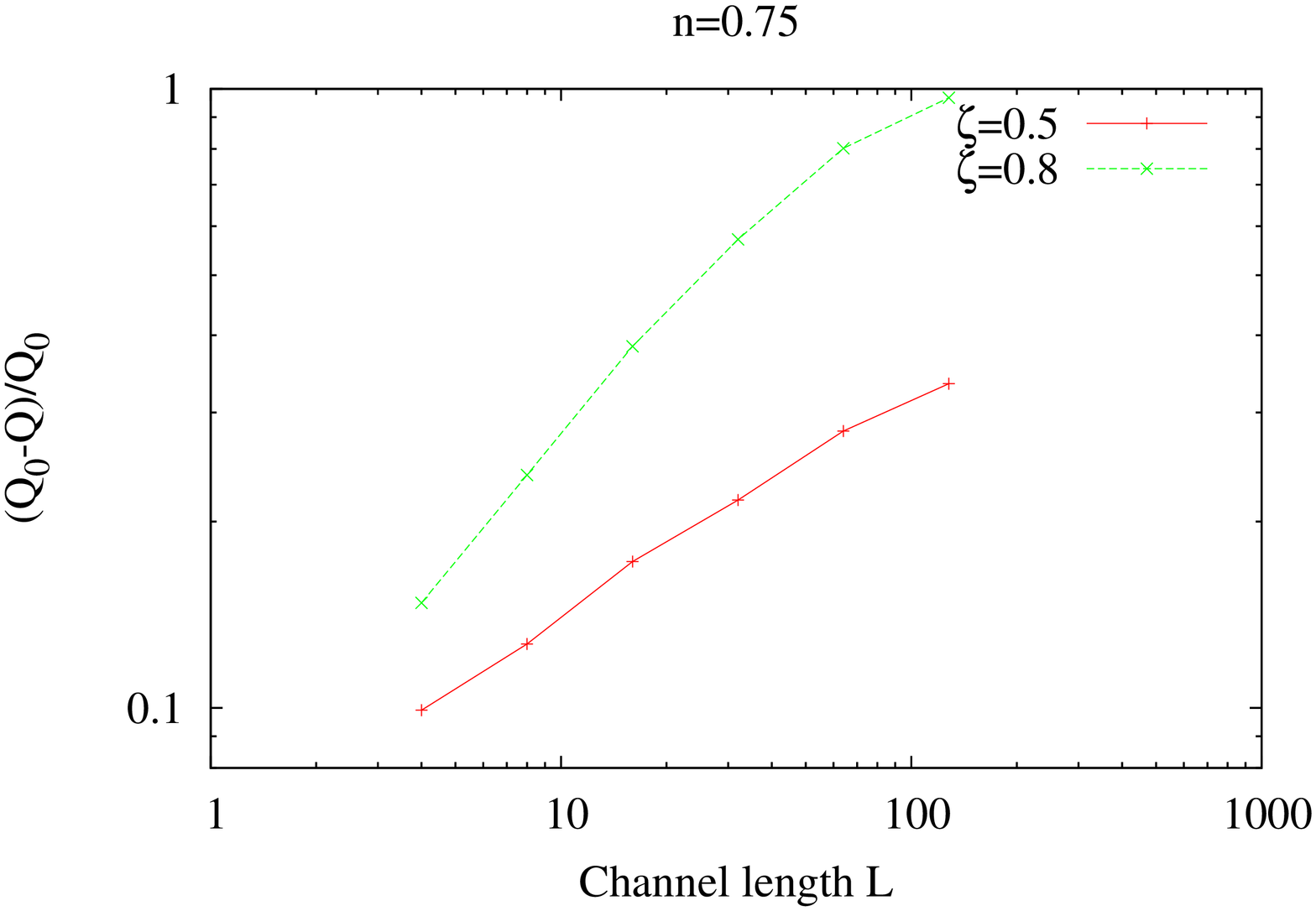}
\caption{Flux variation with length in a channel with one self-affine and one flat
wall for different fluids and Hurst exponents.  The maximum aperture of the channel
is $H_{\rm max}=64$, and the applied pressure gradient is $G=1.0\times 10^{-6}$. Top:
flux depletion for different fluids confined in a channel with $\zeta=0.8$. Right:
flux depletion for a shear-thinning fluid in channels of different $\zeta$.
\label{fig:pw-saff-p}}
\end{center}
\end{figure}

We now consider how the geometry of the fracture effects the (low-Reynolds number)
permeability for the various fluids considered.
First we investigate the effect of the Hurst exponent $\zeta$ on the permeability,
and to simplify the analysis we consider
a fracture channel with one self-affine wall and one flat wall, as in
\cite{drazer2000:permeability}. For a fixed presure gradient $G$, we compute the flux
as a function of the channel length $L$ for the three fluids, and
in Fig.~\ref{fig:pw-saff-p}, we first show the flow rate depletion $(Q_0-Q)/Q_0$
{\em vs}. $L$ for a fracture with $\zeta=0.8$.  Here $Q_0$ is the flux through a
flat-walled channel fo the same mean aperture.  Increasing the length allows for more
fluctuation in the channel width (see Eq.~(\ref{eq:cf}) which increases the tortuosity
and tends to decrease the flux.  If the Hurst exponent of the channel's rough wall is
instead $\zeta=0.5$, the three fluids again behave quite similarly, so it suffices to
compare the behavior of different Hurst exponents for a single case,
and in the lower panel of the figure we plot the flux depletions for the two exponents
for the shear-thinning case.  The fact that the flux depletion is greater for the
$\zeta=0.8$ channel may be explained by noting that this exponent value corresponds to
more fluctuation as a function of $L$ than the 0.5 case, and therefore to a more
tortuous channel.

To relate the flux to the channel aperture, we imagine dividing the channel into a
sequence of nearly-straight sections, each of length $\ell_i$,
and writing the total pressure difference as the sum of the pressure drops in each
section, using Eq.~\ref{eq:flux} for each.  This reasoning yields
\begin{equation}
 \Delta P=\sum _i \Delta P_i=\sum_i \left[ Q \cdot \frac{2n+1}{n} \cdot b_i^{-
\frac{2n+1}{n}} \right]^{n} \cdot m \cdot 2^{n+1} \cdot l_{i},
 \label{eqn:dpsegment}
\end{equation}
where the summation is over the sections, and $b_i$ is the effective aperture and
$l_i$ is the length along the local flow direction in
section $i$, and we have noted that $Q$ is a the same in all sections.
If $\theta_i$ is the angle between the orientation of channel section $i$ and the mean
flow direction, then $b_i=H \cos \theta_i$ and
$l_i=l_i^\Vert/\cos \theta_i$, where $H$ is the aperture and $l^\Vert$ is the
projected length of section $i$ in the mean flow direction, assumed to be the same for
all sections.  Using these relations in Eq.\ref{eqn:dpsegment} we have
\begin{equation}
 \Delta P=2mQ\,l^\Vert\left[2 \frac{2n+1}{n} H^{-\frac{2n+1}{n}} \right]^{n}
\sum_i (\cos \theta_i)^{-(2n+2)}.
 \label{eqn:dpseg1}
\end{equation}
This result generalizes
Eq.~26 in \cite{drazer2000:permeability} to power-law fluids, and if we proceed as in
that reference to evaluate the average over angles $\theta_i$ we obtain
\begin{equation}
 Q-Q_0 \sim H^{(2\zeta-2)/\zeta+(2n+1)/n}
 \label{eqn:qdiff}
\end{equation}
where again $Q_0$ is the flux in a flat channel of the same aperture $H$.

\begin{figure}[htb]
\begin{center}
\includegraphics[scale=.3,angle=-90.]{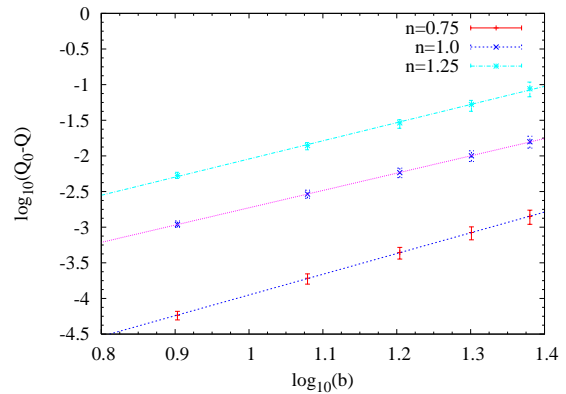}
\caption{Log-log plot of flow rate variations versus the aperture of self-affine
channel for different fluids with power $n=0.75,1.0$ and $1.25$.
\label{fig:saff-law}}
\end{center}
\end{figure}

To test the relation \ref{eqn:qdiff},
we calculate the flow for fracture channels of length $L=256$ with
varying apertures $H=8,12,16,20,24$, for fluid with $m=0.01$, $n$=0.75, 1.0 and 1.25,
all at a pressure gradient $\Delta P/L=1.0e-6$. Figure~\ref{fig:saff-law} shows the
flux depeltion $(Q_0-Q)$ as a function of aperture.  The points are the numerical
results and the solid lines are fitted curves, based on the expected power-law
exponents obtained from Eq.\ref{eqn:qdiff}, which are 2.83, 2.5, and 2.3 for the
three fluids.  We see that the theoretical analysis is in excellent agreement with the
data for the shear-thinning and Newtonian fluids ($n$=0.75 and 1.0), but the agreement
is less statisfactory for the shear-thickening fluid, whose numerical exponent is
closer to 2.5.  A possible interpretation is that in the shear-thickening case,
for the same pressure gradient the average velocity is larger than that for the other
fluids, so that fluid inertia comes into play.

\begin{figure}[htb]
\begin{center}
\includegraphics[scale=.3,angle=-90.]{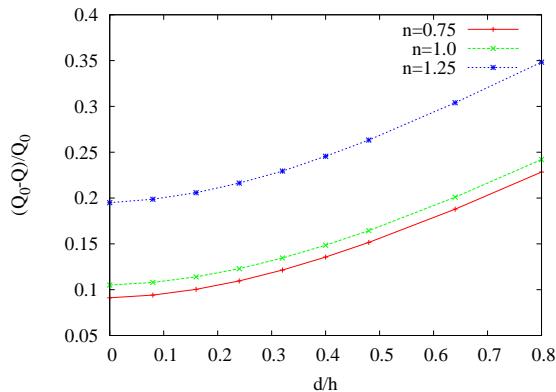}
\caption{Flow rate reduction due to lateral shift along the mean plane of
fracture channel for different fluids with power $n=0.75,1.0$ and $1.25$, and
the Hurst exponent used here is $\zeta=0.8$.
\label{fig:saff-shift}}
\end{center}
\end{figure}

Finally, we consider an additional effect, a lateral shift between the two sides of a
fracture, which might arise in practice due to geological processes.
We begin with a fracture channel with complimentary sides and constant initial
aperture $H$, and then shift one side along the mean plane by a distance $d$.
The fracture aperture is now a function of position, $H_d(x)$, and effectively
a spatial random function.  We again compute the flux depletion relative to a flat
channel having the same initial aperture, using six
realizations of a self-affine fracture wall with Hurst exponent $\zeta=0.8$.
As shown in Fig.~\ref{fig:saff-shift} the flux decreases somewhat faster than
linearly with shift, by producing narrow gaps when proturbances on the two sides are
brought closer to one another.  The shear-thinning and Newtonian fluids have a fairly
similar behavior, while the reduction is twice as large in the shear-thickening case,
perhaps again as a result of inertial effects.  As in the previous discussion, using a
different value $\zeta=0.5$ for the Hurst exponent gives the same trends.

\section{Conclusion}
\label{sec:concl}
Using a new implementation of the lattice Boltzmann method for power-law fluids,
we have investigated their flow in two-dimensional
self-affine fracture channels as a function of applied pressure gradient.
Generally, fluids with different power-law index behave in a similar manner
when their flow parameters are properly scaled, using standard results for flow in
constant-thickness channels.  Many previous results for Newtonian fluids in
self-affince fractures  are found to generalize in a straightforward manner.
However, shear-thickening fluids, which have higher velocities for the same
pressure gradient than Newtonian or shear-thnning counterparts, are more
susceptible to inertial effects.

With regard to the local flow fields, we first considered the maximum absolute velocity
as a function of distance along the mean flow direction, which was found to
fluctuates along the fracture channel due to its tortuosity and the variable
effective aperture along the channel. The local maxima of this maximum
absolute velocity occur at points of narrowing or minimal effective aperture,
and the range of maximum absolute velocity relative to the global mean velocity
ranges from about 1.5 to 5.5.
With increasing inertia, this normalized maximum absolute velocity
increases for all power-law fluids to different degrees, with shear-thickening
fluids having the largest effect and shear-thinning the least.
As the pressure gradient increases, the normalized maximum velocities near
the constrictions are relatively constant but outside these points velocities
tend to increase.  The variation in velocity is greatest for a shear-thickening
fluid and lest for shear-thinning.  Pressure fluctuations along the
channel increase with forcing for all fluids, and for a given pressure gradient
increase with the power-law index $n$.

The relationship between pressure gradient and flux is found to have the same
functional form as for flow in a flat channel, $\Delta p\sim Q^n$, when inertial
effects are absent.  At higher $\Delta p$, Newtonian fluids behave in the same way
as in intergranular porous media, and shear-thinning fluids behave analogously, but
the shear-thickening case does not show simple power-law behavior. It is possible to
collapse all of the data on flux {\em vs}. pressure gradient into a universal
friction factor curve.  The variation of flux with system length was  shown to scale
with system length with an exponent algebraically related to the Hurst exponent,
in a manner which generalizes the Newtonian case.

The most interesting question raised by these results is the form of the flux-pressure
gradient relationship in the regime of strong inertia in the non-Newtonian case.
In this work, we were limited in the range of accessible pressure gradients by
numerical instabiliites, and it is desirable to improve the algorithm so as to
consider higher pressure gradients and further
explore the dynamics of the inertial regime.  An extension of these considerations to
viscoelastic fluids is likewise highly desirable, but new ideas beyond the methods
used here are needed.

\begin{acknowledgments}
This work was supported by the NSF CREST Center for Mesoscopic Modeling and
Simulation at CCNY, and the Geosciences Program of the Office of Basic Energy
Sciences, U. S. Department of Energy. We thank German Drazer and Jean-Pierre
Hulin for helpful discussions and suggestions.
\end{acknowledgments}

\end{document}